\documentclass[eqsecnum]{emulateapj}
\usepackage{natbib,amsmath}
\usepackage{epsfig}
\usepackage{hyperref}
\shorttitle{X-ray spectra of PKS 2155--304}
\shortauthors{Gaur et al.}
\usepackage{color}

\begin{document}

\title{The Hard X-ray emission of the blazar PKS 2155--304 }

\author{Haritma Gaur\altaffilmark{1}, Liang Chen\altaffilmark{1,2}, R. Misra\altaffilmark{3}, S. Sahayanathan\altaffilmark{4},
M. F. Gu\altaffilmark{1}, P. Kushwaha\altaffilmark{5}, G. C. Dewangan\altaffilmark{3} }
\altaffiltext{1}{Key Laboratory for Research in Galaxies and Cosmology, Shanghai Astronomical Observatory,
Chinese Academy of Sciences, 80 Nandan Road, Shanghai 200030, China; haritma@shao.ac.cn}
\altaffiltext{2}{University of Chinese Academy of Science, 19A Yuquanlu, Beijing 100049, China; chenliang@shao.ac.cn}
\altaffiltext{3}{Inter-University Centre of Astronomy and Astrophysics (IUCAA), India}
\altaffiltext{4}{Astrophysical Sciences Divison, Bhabha Atomic Research Center, Mumbai, India}
\altaffiltext{5}{Department of Astronomy (IAG-USP), University of Sao Paulo, Sao Paulo 05508-090, Brazil}
\begin{abstract}

The synchrotron peak of the X-ray bright High Energy Peaked Blazar (HBL) PKS 2155$-$304 occurs in the UV-EUV region and hence its
 X-ray emission (0.6--10 keV) lies mostly in the falling part of the synchrotron hump.
We aim to study the X-ray emission of PKS 2155$-$304 during different intensity states in 2009$-$2014 using XMM$-$Newton satellite.
We studied the spectral curvature of all of the observations to provide crucial information on the energy distribution of the
non-thermal particles. Most of the observations show curvature or deviation from a single power-law and can be well modeled by a
log parabola model. In some of the observations, we find spectral flattening after 6 keV. In order to find the possible
origin of the X-ray excess, we built the Multi-band Spectral Energy distribution (SED). We find that the X-ray excess in PKS 2155--304
is difficult to fit in the one zone model but, could be easily reconciled in the spine/layer jet structure.
 The hard X-ray excess can be explained by the inverse Comptonization of the synchrotron photons (from the layer) by
the spine electrons. 
\end{abstract}


\section{Introduction}

Blazars are multi-wavelength, multi-timescale phenomena (Ulrich et al. 1997) and their  extreme properties are thought to be due
to relativistic jets  pointing nearly to our line of sight ($\la$ 10$^{\circ}$)  (e.g., Urry \& Padovani 1995).
Blazar light curves show an erratic and unpredictable behaviour in terms of flare strength and duration from event to event.

The broad band (from IR to TeV energies) spectral energy distribution (SED) of blazars are characterized by a broad double peaked structure.
The peak of the low energy spectral component is found at X-ray/UV energies for  High energy Peaked Blazars (HBLs) and at optical energies
for Low energy peaked Blazars (LBLs). The peak of the high energy spectral component for HBLs occurs at GeV/TeV energies while for LBLs it
is usually at MeV/GeV energies. In fact, the peak of the first spectral component is known to inversely correlate with the
 luminosity of the blazar and
different kind of blazars can be classified based on their peak energy to form a sequence referred to as the blazar sequence
 (Fossati et al. 1998; Ghisellini et al. 1998; Giommi et al. 2012).

Phenomenologically the SED of TeV blazars can be roughly explained by invoking
non-thermal electrons which are approximately distributed in
energy in  a broken power law shape (Fossati et al. 2000a, b, Sauge et al. 2006). The low energy spectral
component of the SED arises from the synchrotron emission of these
particles. The high energy component is then modeled as synchrotron
self Compton and/or external photon inverse Comptonization by the same non-thermal
distribution. The non-thermal particles are believed to be produced in a so called  ``acceleration'' region,
where the balance of the acceleration and escape time-scales, produces a single
power-law distribution. These electrons escape into a larger ``cooling'' region where
they produce the observed synchrotron and inverse Compton components (Kirk, Rieger \& Mastichiadis 1998).
Since the radiative cooling rate of these processes is proportional to
the energy square of the electrons, the high energy electrons are affected
by the cooling, while the low energy ones are not. This differentiates the
electron population into two regimes and it is believed that the two power-law
forms of the empirical broken power-law correspond to these two regimes (Pacholczyk 1970; Kardashev 1962).
While this interpretation is popular and standard, it should be noted that
theoretically the spectral index difference between the two power-laws
predicted from such a radiative
cooling model should be $\sim 0.5$ which is not often observed (Sikora et al. 2009 and references therein).

A simple description of the X-ray data of blazars is the log parabola model. Here the power-law index is not a constant but
varies slowly with energy i.e. $\propto$ log E and hence the name
log parabola. This model has often been invoked to fit the entire SED of blazars
(Landau et al. 1986; Massaro et al. 2004; Chen et al. 2014) and such curved spectra of blazars are known to arise by
synchrotron or inverse Compton radiation from electron distributions featuring in log parabola shape (Tramacere et al. 2007; Paggi et al. 2009).
There is a correlation observed between the curvature parameter of the log parabola
model $\beta$ and the energy at which the component peaks
(Massaro et al. 2004; 2008; 2011a; Tramacere et al. 2007;2009) which is consistent with the
theoretical expectations (Tramacere et al. 2007; Paggi et al. 2009).
However, it is unlikely that a single log parabola model is the true representative of the complete SED data and indeed for several
optical--X-ray SED fits it is found to be inadequate, requiring additional components (e.g. Bhagwan et. al. 2014). Massaro et al. (2004) and



PKS 2155--304 is the brightest X-ray BL Lac (Griffiths et al. 1979; Schwartz et al. 1979;  Brinkmann
et al. 1994; Giommi et al. 1998) and is a confirmed TeV $\gamma$-ray
source by the observations of H.E.S.S. collaboration in 2002 and 2003
(Aharonian et al. 2005). The synchrotron hump peaks in the UV/EUV and is $< 100$ eV even in the very high states
 (Zhang et al. 2002; Massaro et al. 2008). The brightness of the source and that it
has been observed several times by X-ray satellites makes it an
excellent source to study its spectral curvature. It has been an important target of various X-ray satellites like ASCA,
BeppoSAX,  EGRET, RXTE, HESS, Swift XRT, XMM-Newton (Sreekumar \&
Vestrand 1997; Vestrand \& Sreekumar 1999; Tanihata et al.\ 2001;
Zhang et al.\ 2002; 2005; Massaro et al.\ 2004;  Aharonian et
al.\ 2005; Foschini et al.\ 2007; H.E.S.S.~Collaboration et al. 2014; Kapanadze et al. 2014). Depending on various flux states,
several works have reported
curvature in the X-ray band (Fossati et al. 2000a, Foschini et al. 2007, Massaro et al. 2008, Chen et al. 2014
and references therein).  On several occasions, the $\gamma$-ray
luminosity of PKS 2155--304 dominates over the low energy synchrotron
component, a trend similar to flat spectrum radio quasars
(e.g. Massaro et al.\ 2011a; Aharonian et al.\ 2009; Zhang et al.\ 2008) .

The large effective area and high spectral resolution of {\it XMM-Newton} has provided unprecedented X-ray spectral information
for PKS 2155--304. {\it XMM-Newton} has been observing the source frequently since 2000. In a sample study of several blazars,
Massaro et al. (2008) analyzed a number of {\it XMM-Newton} observations of  PKS 2155--304 before 2007 and fit all of them
using log parabolic model.
They found that the curvature versus peak energy correlation was similar to other blazars while the curvature
versus peak luminosity was not. On the other hand, Zhang (2008) reported spectral hardening (i.e. negative curvature) for two
observations in 2006. The spectral curvature is very small (i.e. -0.05 to -0.09) and interpreted it as the detection of
the Inverse Compton component in the X-ray band. However, most observations of the source show regular positive curvature
and the above two observations may be taken as special cases when the inverse component was revealed. 
Bhagwan et al. (2014) analyzed 20 XMM--Newton observations of PKS 2155--304 during the period 2000--2012 and model
the simultaneous Optical/UV--X-ray SED usind the log-parabolic model. Four observations from 2009--2012 which we are analyzing 
in this work are presented by them. They found that log-parabolic model is not adequate and
require additional power law component to model the simultaneous Optical/UV--X-ray SED. Recently, using
NuSTAR observations, Madjeski et al. 2016 found significant hard X-ray excess for PKS 2155--304 during epoch 2013 amd model the 
Multiband SED with the one zone model.

Here, our motivation is to study the X-ray spectral curvature of PKS 2155--304 during the period 2009--2014
using the XMM$-$Newton observations.
We study the X-ray spectra of PKS 2155--304 during various flux states and tested if they are well described
with single power-law or curved log parabola model. In some of the observations, we found concave X-ray spectra/X-ray excess.
In order to find the possible physical origin of concave X-ray spectra, we construct the simultaneous Multi-wavelength SEDs of 
PKS 2155--304 and determine various parameters of the blazar jet.

The paper is structured as follows. In Section 2, we give a brief description of the Observations and the data reduction method.
 The results of the analysis are presented in section 3 which are discussed and concluded in Section 4.

\begin{table}
{\caption{Observation log of PKS 2155--304 with XMM-Newton PN}}
\textwidth=6.0in
\scriptsize
\setlength{\tabcolsep}{0.035in}
\noindent
\begin{tabular}{cccc} \hline \hline

Date of Observation   & Observation & GTI $^a$(ks)   &$F_{var}$ \\
 dd.mm.yyyy           &ID           &(in Ks)        &(0.6--10) keV \\\hline
28.05.2009  &0411780401  & 64820   &9.17$\pm$0.08    \\
28.04.2010  &0411780501  & 60900   &10.14$\pm$0.12 \\
26.04.2011  &0411780601  & 63818   &7.02$\pm$0.09  \\
28.04.2012  &0411780701  & 68735   &4.00$\pm$0.27  \\
23.04.2013  &0411782101  & 66300   &5.55$\pm$0.16   \\
25.04.2014  &0727770901  & 61000   &3.40$\pm$0.14  \\ \hline
\end{tabular}

$^a$ GTI (Good Time Interval).
\label{Tabobs}
\end{table}

\begin{table}
{\caption{Observation log of PKS 2155--304 with XMM-Newton OM (Imaging Mode)}}
\textwidth=6.0in
\scriptsize
\setlength{\tabcolsep}{0.035in}
\noindent
\begin{tabular}{ccccc} \hline \hline

Date of Observation   &Filter &Count  &Magnitude$^a$  &Flux$^b$  \\
 dd.mm.yyyy           &       &Rate    &           &  \\ \hline
28.05.2009            &V      &110.13$\pm$0.58 &12.86$\pm$0.01 &2.75$\pm$0.01   \\
                      &U      &256.13$\pm$0.54 &12.24$\pm$0.01 &4.97$\pm$0.01\\
                      &B      &266.20$\pm$0.06 &13.20$\pm$0.01 &3.33$\pm$0.01 \\
                      &UVW1   &119.91$\pm$0.18 &12.01$\pm$0.01 &5.78$\pm$0.01 \\
                      &UVM2   &35.08$\pm$0.12  &11.91$\pm$0.01 &7.75$\pm$0.03 \\
                      &UVW2   &14.16$\pm$0.07  &11.99$\pm$0.01 &8.07$\pm$0.04  \\ \hline
28.04.2010            &V      &56.84$\pm$0.35  &13.58$\pm$0.01 &1.42$\pm$0.01  \\
                      &U      &124.92$\pm$0.68 &13.02$\pm$0.01 &2.42$\pm$0.01 \\
                      &B      &133.73$\pm$0.71 &13.95$\pm$0.01 &1.67$\pm$0.01 \\
                      &UVW1$^c$&62.15$\pm$0.14 &12.72$\pm$0.01 &3.00$\pm$0.01 \\
                      &UVM2   &18.95$\pm$0.09  &12.58$\pm$0.01 &4.19$\pm$0.02 \\
                      &UVW2   &7.81$\pm$0.05   &12.63$\pm$0.01 &4.45$\pm$0.03 \\ \hline
26.04.2011            &V      &67.30$\pm$0.34  &13.39$\pm$0.01 &1.68$\pm$0.01   \\
                      &U      &146.81$\pm$0.29 &12.84$\pm$0.01 &2.85$\pm$0.01 \\
                      &B      &157.66$\pm$0.31 &13.77$\pm$0.01 &1.97$\pm$0.01 \\
                      &UVW1   &68.81$\pm$0.02  &12.61$\pm$0.01 &3.32$\pm$0.01 \\
                      &UVM2   &20.05$\pm$0.09  &12.52$\pm$0.01 &4.43$\pm$0.02 \\
                      &UVW2   &8.37$\pm$0.06   &12.56$\pm$0.01 &4.77$\pm$0.03 \\ \hline
28.04.2012            &V      &31.26$\pm$0.17  &14.23$\pm$0.01 &0.78$\pm$0.01  \\
                      &U      &64.68$\pm$0.28  &13.73$\pm$0.01 &1.25$\pm$0.01 \\
                      &B      &71.24$\pm$0.33  &14.63$\pm$0.01 &0.89$\pm$0.01 \\
                      &UVW1   &31.07$\pm$0.01  &13.47$\pm$0.01 &1.50$\pm$0.01 \\
                      &UVM2   &9.25$\pm$0.06   &13.36$\pm$0.01 &2.04$\pm$0.01 \\
                      &UVW2   &3.67$\pm$0.04   &13.46$\pm$0.01 &2.09$\pm$0.02 \\ \hline
23.04.2013            &V      &56.10$\pm$0.26  &13.59$\pm$0.01 &1.40$\pm$0.01  \\
                      &U      &118.43$\pm$0.24 &13.08$\pm$0.01 &2.30$\pm$0.01 \\
                      &B      &126.29$\pm$0.25 &14.01$\pm$0.01 &1.58$\pm$0.01 \\
                      &UVW1   &57.19$\pm$0.19  &12.81$\pm$0.01 &2.76$\pm$0.01 \\
                      &UVM2   &16.64$\pm$0.08  &12.72$\pm$0.01 &3.68$\pm$0.02 \\
                      &UVW2   &6.93$\pm$0.05   &12.77$\pm$0.01 &3.95$\pm$0.03 \\\hline
25.04.2014            &V      &61.55$\pm$0.03  &13.49$\pm$0.01 &1.54$\pm$0.01 \\
                      &U      &131.50$\pm$0.25 &12.96$\pm$0.01 &2.55$\pm$0.01 \\
                      &B      &139.05$\pm$0.26 &13.91$\pm$0.01 &1.74$\pm$0.01 \\
                      &UVW1   &63.45$\pm$0.21  &12.70$\pm$0.01 &3.06$\pm$0.01 \\
                      &UVM2   &18.38$\pm$0.09  &12.61$\pm$0.01 &4.06$\pm$0.02 \\
                      &UVW2   &7.57$\pm$0.05   &12.67$\pm$0.01 &4.31$\pm$0.03 \\ \hline
\end{tabular}

$^a$ Instrumental Magnitude. \\
$^b$ Flux in units of 10$^{-14}$ erg cm$^{-2}$ s$^{-1}$ A$^{-1}$. \\
\end{table}

\begin{table*}
\textwidth=6.0in
\scriptsize
{\caption {Best fit spectral parameters for the power law, log parabolic and broken power law model for the whole observation.} }
\noindent
\begin{tabular}{ccccccccc} \hline \hline

Year   &$\Gamma_{1}$  &E$_{b}$    &log$_{10}$Flux  &$\Delta \Gamma$/$b$  &$\chi_{Red}^{2}$  &dof &F-test &P-value\\
Segment&              & (keV)               &                  &                &                  &  && \\ \hline
2009 &$2.838_{-0.003}^{+0.003}$  &-                           &$-10.117_{-0.001 }^{+0.001}$ &-    &2.98 &164 &- &-   \\
2010 &$2.831_{-0.004}^{+0.004}$  &-                           &$-10.399_{-0.001 }^{+0.001}$ &-    &1.11 &1077 &- &- \\
2011 &$2.572_{-0.003}^{+0.003}$  &-                           &$-10.141_{-0.001 }^{+0.001}$ &-    & 2.25&165 &- &- \\
2012 &$2.867_{-0.008}^{+0.008}$  &-                           &$-10.822_{-0.001 }^{+0.001}$ &-    & 1.40&155 &- &- \\
2013 &$2.787_{-0.002}^{+0.002}$  &-                           &$-10.449_{-0.001 }^{+0.001}$ &-    &1.92 &159 &- &- \\
2014 &$2.875_{-0.004}^{+0.004}$  &-                           &$-10.445_{-0.001 }^{+0.001}$ &-    &2.27 &156 &- &-\\ \hline

2009 &$2.767_{-0.008 }^{+0.008}$ &- &$-10.120_{-0.001 }^{+0.001}$ &$0.093_{-0.010 }^{+0.010}$ &1.50 &163 &160.83 &4.44$\times10^{-26}$\\
2010 &$2.768_{-0.012 }^{+0.011}$ &- &$-10.402_{-0.001 }^{+0.001}$ &$0.082_{-0.014 }^{+0.014}$ &1.03 &1076 &83.57 &2.98$\times10^{-19}$\\
2011 &$2.504_{-0.008 }^{+0.008}$ &- &$-10.145_{-0.001 }^{+0.001}$ &$0.082_{-0.009 }^{+0.009}$ &1.00 & 164 &205.00 &1.10$\times10^{-30}$\\
2012 &$2.916_{-0.024 }^{+0.024}$ &- &$-10.818_{-0.003 }^{+0.003}$ &$0.070_{-0.032 }^{+0.032}$ &1.33 &154  &8.11  &0.005\\
2013 &$2.728_{-0.013 }^{+0.013}$ &- &$-10.452_{-0.001 }^{+0.001}$ &$0.074_{-0.015 }^{+0.015}$ &1.50 & 158 &44.24 &4.50$\times10^{-10}$\\
2014 &$2.790_{-0.013 }^{+0.013}$ &- &$-10.449_{-0.001 }^{+0.001}$ &$0.108_{-0.015 }^{+0.015}$ &1.40 & 155 &96.32 &5.53$\times10^{-18}$\\  \hline

2009 &$2.812_{-0.004 }^{+0.004}$ &$2.123_{-0.134 }^{+0.157}$  &$-10.120_{-0.001 }^{+0.001}$ &$  0.119_{-0.014 }^{+0.016}$ &1.44 & 162 &- &-\\
2010 &$2.789_{-0.014 }^{+0.010}$ &$1.342_{-0.161 }^{+0.123}$  &$-10.402_{-0.001 }^{+0.001}$ &$  0.084_{-0.014 }^{+0.014}$ &1.03 &1075 &- &-\\
2011 &$2.546_{-0.004 }^{+0.004}$ &$2.359_{-0.164 }^{+0.198}$  &$-10.145_{-0.001 }^{+0.001}$ &$  0.114_{-0.015 }^{+0.016}$ &0.89 & 163 &- &-\\
2012 &$2.871_{-0.008 }^{+0.009}$ &$6.259_{-0.427 }^{+0.545}$  &$-10.808_{-0.004 }^{+0.004}$ &$ -2.167_{-1.011 }^{+0.834}$ &1.16 & 153 &- &-\\
2013 &$2.765_{-0.006 }^{+0.006}$ &$2.217_{-0.165 }^{+0.154}$  &$-10.452_{-0.001 }^{+0.001}$ &$  0.103_{-0.020 }^{+0.020}$ &1.42 & 157 &- &-\\
2014 &$2.844_{-0.010 }^{+0.006}$ &$2.129_{-0.455 }^{+0.164}$  &$-10.449_{-0.001 }^{+0.001}$ &$  0.135_{-0.029 }^{+0.023}$ &1.36 & 154 &- &-\\ \hline

\end{tabular} \\
\noindent{$\Gamma_1$: Low energy spectral index; $E_b$: Break Energy; $b$: curvature; Flux in ergs/sec/cm$^2$; $\Delta \Gamma$:
Difference of high and low energy spectral index; $\chi_{Red}^{2}$: Reduced $\chi^2$; dof: degree of freedom }
\end{table*}

\section{XMM--Newton Observations and Data Analysis}

\subsection{X-ray Data}

PKS 2155--304 is observed by the European Photon Imaging Camera (EPIC)
on board the XMM-Newton satellite (Jansen et al.\ 2001). We considered six
observations that took place between May 2009--April 2014 with roughly a difference of an year.
The observation log is given in Table \ref{Tabobs}.
The EPIC is composed of three co-aligned X-ray telescopes  which
simultaneously observe a source by accumulating photons in the three CCD-based instruments:
the twins MOS 1 and MOS 2 and the pn (Turner et al.\ 2001; Str\"{u}der et al.\ 2001).
The EPIC instrument provides imaging and spectroscopy in the energy range from 0.2 to
15 keV with a good angular resolution (PSF = 6 arcsec FWHM) and a moderate spectral
resolution ($E/\Delta E \approx 20-50$). We consider here only the EPIC-pn data as it is most sensitive and
 less affected by the photon pile-up effects.

We used the XMM-Newton Science Analysis System (SAS) version 15.0.0 for the light curve extraction
and spectral analysis. The summary file of the Observation Data File (ODF) and the calibration
index file (CIF) are generated using updated calibration data files or Current Calibration
Files (CCF) following "The XMM$-$Newton ABC Guide" (version 4.6, Snowden et al. 2013).
 In all of our observations, EPIC-pn detector was operated in small window (SW) imaging mode.
XMM$-$Newton EPCHAIN pipeline is used to generate the event files.
In order to identify intervals of flaring particle background, we extracted the  high energy (10 keV $< E <$ 12 keV)
light curve for the full frame of the exposed CCD and no significant background flares were
found. Pile up effects are examined for each observation by using the SAS task EPATPLOT and we found that mostly triple and quadruple
events are affected by the pile-up effects hence, we extracted only the single and double events for our observations.
In order to further check the pile up issue, we extracted the spectra by excluding a circular region with a radius of 10 arcsec
centered on the source for our observations. We again calculated the best fit parameters of power law and log parabola model
on these spectra and found the parameters to be consistent within the error bars with our fitting results obtained without
excluding the central region.
We read out source photons recorded in the entire 0.6 $-$ 10 keV energy band, using a circle of 45 arcsec
radius centered on the source. Background photons were read out from a circular region with
an area comparable to the source region, located about 180 arcsec off the source but
on the same chip set.
Redistribution matrices and ancillary response files were produced using the SAS tasks {\it  rmfgen} and
{\it arfgen}. The pn spectra were created by the SAS tool {\it XMMSELECT} and grouped to have at least 30 counts
in each energy bin to ensure the validity of $\chi^{2}$ statistics.

\subsection{Optical/UV data}

The Optical/UV MonitorTelescope (OM hereafter) onboard XMM$-$Newton provides the facility to observe a source simultaneously in optical
and UV bands along with the X-ray bands, with a very high imaging sensitivity (Mason et al. 2001).
The OM can collect data with time resolution of 0.5 s for the wavelength range 170$-$650 nm with six broad-band filters, three in the
optical and three in the UV, with an FOV covering 17 arcmin of the central region of the X-ray telescope FOV.
The optical U, B and V filters collect data in the wavelength ranges 300$-$390 and 390$-$490, 510$-$580 nm respectively, and the ultraviolet
UV W2, UV M2 and UV W1 filters collect data in the wavelength ranges 180$-$225, 205$-$245 and 245$-$320 nm, respectively. PKS 2155$-$304
was observed with OM in imaging mode in all of the filters during these observations. We reprocess the OM imaging data using the standard
{\it omichain} pipeline of SAS, from which we get a {\it combolist} file containing the calibrated data with their errors for all
the sources which are present in the field of view.The counts, instrumental magnitude and the fluxes for PKS 2155$-$304 are extracted
from the file and are provided in Table 2.

\section{Results}

\subsection{Temporal Variability}
The XMM--Newton optical/UV/X-ray light curves of PKS 2155$-$304 are presented in Figure 2. Each observational point in the figure
indicates the average count rate in the X-ray (0.6--10 keV) band and OM filters.
It can be seen that the source is variable in all of the bands during 2009-2014. The fractional variability amplitudes corresponding
to the X-ray, UVW1, UVM2, UVW2, U,B and V bands are 55.12$\pm$1.13, 43.33$\pm$0.09, 42.87$\pm$0.19, 42.18$\pm$0.27, 45.01$\pm$0.12,
43.18$\pm$0.12 and 40.41$\pm$0.21, respectively. The variability amplitudes at optical and UV frequencies are smaller than that observed at X-ray frequencies
and is consistent with the previous findings (Zhang et al. 2006; Ostermann et al. 2007). The correlation between X-ray versus optical/UV and
optical versus UV are shown in fig 3. It can be seen that optical and UV emission are significantly correlated (Pearson correlation coefficient,
$r$=0.99, $p$=5.96$\times 10^{-6}$) whereas there is weak correlation between X-ray w.r.t optical/UV (Spearman correlation coefficient,
$\rho$=0.89, $p$= 0.02).

During all the six observation epochs, X-ray variability is clearly detected with recurrent flare like events.

\subsection{Spectral Variability}

The XSPEC software package version 12.8.1 is used for spectral fitting. The Galactic absorption  $n_{H}$ was fixed to 1.52
$\times$ 10$^{20}$ cm $^{-2}$ (Lockman \& Savage 1995) and the Xspec routine `cflux' was used to obtain unabsorbed flux and its error.

 The spectra are fitted with a simple power law defined as kE$^{\Gamma}$, a broken power law which is defined as kE$^{\Gamma_{1}}$
for E $<$ $E_{break}$ ; kE$^{\Gamma_{2}}$ otherwise and a log parabolic model.
In the log parabolic model, the spectrum is assumed to be $\propto E^{-\Gamma (E)}$ where the
photon index $\Gamma$ is not a constant but varies logarithmically with energy i.e.
\begin{equation}
\Gamma (E) = \alpha + \beta log(E/E_{1})
\label{spec}
\end{equation}
Here, $\alpha$ is a parameter denoting the local spectral index at $E_1$ and $\beta$ is the
curvature parameter. Together with the normalization, $\alpha$ and $\beta$ are the three
parameters of the spectrum. Without loss of generality, $E_1$ is fixed to some convenient value which
in our case is the lowest energy of the spectra $E_{1} = 0.6$ keV.
 The spectral parameters of all the models are presented in Table 3.
In most of the observations, log parabolic models give better descriptions by giving systematically lower $\chi^{2}$ values
which can be seen by the F-values in Table 3. The spectral fitting for all of the observations are shown in fig 4.

In 2012 pointing, it can be seen that the residuals around 6 keV shows deviations in the upward direction.
The spectrum shows a break at around $E_{p}$ $\sim$ 6.3 keV indicating the flattening in the spectrum.
The log parabolic model also shows mild negative curvature of -0.07. It can be noted that this pointing has the lowest flux among all
of the observations. From the spectral fitting of other observations, it can be seen that there are positive residuals around 7 keV.
Although, the fitting parameters do not show any negative curvature for these observations.
The spectral flattening/negative curvature could be attributed to the weak contamination of the IC component in the
synchrotron component as reported in previous studies (e.g., Zhang 2008; Madejski et al. 2016). In order to further
search for the physical origin of these positive residuals, we contruct the multi wavelength spectra of these six epochs.

\subsection{Multi-wavelength Spectral Energy Distribution}

As discussed above, XMM--Newton observation of PKS 2155--304 during 2012  shows significant concave X-ray spectra. Fortunately,
Nuclear Spectroscopic Telescope Array ({\it NuSTAR}) satellite has simultaneous observation of PKS 2155--304 in 2013 with our
observations of XMM--Nweton.
This data of PKS 2155--304 has been successfully reduced by \citet{2016ApJ...831..142M} and they found significant X-ray spectral
flattening during epoch 2013. Hence, we also included the {\it NuSTAR} data of this epoch in our SED fitting.
 To explore the origin of such concave feature, we construct simultaneous broadband SEDs from radio through $\gamma$-ray for
our observations from epochs 2009-2014. {\it Fermi}/LAT $\gamma$-ray data and radio data reductions are provided below in Subsections
\ref{sec:sed_gamma-ray} and \ref{sec:sed_radio}, respectively. PKS 2155-304 is a TeV $\gamma$-ray source
which is detected by the H.E.S.S. collaboration \citep{2005A&A...430..865A}. Unfortunately, we do not have
simultaneous TeV data corresponding to our observations. Hence, we collect the historical TeV spectra from \citet{2012A&A...539A.149H}
just to guide the broadband SEDs. The TeV photons can be absorbed by
extragalactic background light (EBL) through pair production effect. Therefore, we correct the observed TeV spectra by
employing the EBL model of \citet{2008A&A...487..837F}. For the epoch 2013, we collect the hard X-ray data from
\citet{2016ApJ...831..142M} (mainly detected by {\it NuSTAR}) to make the concave X-ray spectra more clear.
Finally, out of six simultaneous broadband SEDs (except TeV band), two X-ray spectra have shown concave feature , i.e.,
during epochs 2012 and 2013. All these 6 broadband SEDs are presented in Figure \ref{fig:sed}.

\subsubsection{$\gamma$-ray Data}
\label{sec:sed_gamma-ray}

Fermi Large Area Telescope (LAT) is a gamma-ray imaging telescope, continuously scanning the whole sky every $\sim$3 hours for
photons with energies  $>20$ MeV \citep{2009ApJ...697.1071A}. The data used here for generating the gamma-ray SEDs are of 1
month duration centered on the corresponding XMM observations and were analysed following the standard procedure for PASS 8
instrument response function. The 100 MeV--300 GeV energy bin was divided into multiple bins and in each, only
photon like events classified with keywords ‘ EVCLASS =128, EVTYPE =3’ were selected from a
15$^{\circ}$ circular region centered on the source position. The gamma-ray events from the Earth's limb
were minimized by restricting the events to a maximum zenith angle of 90$^{\circ}$. These events were
then further filtered with the instrument good time intervals using the standard cut
"( DATA\_QUAL $>$0) \&\&(LAT\_CONFIG==1)". The resulting events were then modeled using the
maximum likelihood method with an input model provided file and associated exposure map calculated
on ROI and additional 10$^{\circ}$ around it. The source model file was generated from the 3rd Fermi-LAT catalog (3FGL –
gll\_psc\_v16.fit; Acero et al. 2015) including the Galactic (gll\_iem\_v06.fits) and  extra-galactic
(iso\_P8R2\_SOURCE\_V6\_v06.txt) contribution through the respective template provided by the
LAT team.

For each energy bin, the fitting was performed iteratively, removing insignificant source with Test Statistics $<0$
following \citet{2014ApJ...796...61K} until it converged. For PKS 2155-304, we used a power-law model.
The best fit value was finally used to derive the energy flux to construct the gamma-ray SED.

\subsubsection{Radio Data}
\label{sec:sed_radio}

Radio observations at 22.2 GHz are carried out with the 22-m radio telescope (RT-22) at the Cromean Astronomical
Observatory (CrAO). Observations at 37.0 GHz are made with the 14 m radio telescope of Aalto University’s Metsahovi Radio
Observatory in Finland. For our measurements, we used two similar Dicke switched radiometers of 22.2 and 36.8 GHz.
More details about the observations and data reduction are provided in Gaur et al. (2015).

\subsubsection{One zone model}
\label{sec:sed_one_zone_model}

The multiwavelength emission of HSP BL Lacs often show correlated variability and therefore are thought to be produced from a
single emitting region, except the radio emission, which is probably the superimposed emissions of many optically
thick emitting regions along the jet \citep[e.g.,][]{1979ApJ...232...34B, 1985A&A...146..204G}. One zone synchrotron + synchrotron
self Compton (SSC) model is successful in explaining the multi-wavelength SED of HSP BL Lacs (except radio band). Therefore, we adopt
the one zone model to fit the SEDs of PKS 2155-304, and the concave X-ray spectra is thought to the composite spectra of synchrotron
and SSC emissions produced from same electron population as also reported by Zhang (2008) and Madejski et al. (2016).
Since, we are interested in those SEDs which have concave X-ray spectra, therefore in this work we only model 
two SEDs of epochs 2012 and 2013. We briefly describe the model here. For detailed information, please see
\citet{2017ApJ...842..129C}. The model assumes a homogeneous and isotropic emission region,
which is a sphere with radius $R$ that has a uniform magnetic field with strength $B$ and a uniform electron energy
distribution as described by,

\begin{equation}
N(\gamma)=\left\{ \begin{array}{ll}
                    N_{0}\gamma ^{-p}  &  \mbox{ $\gamma_{\rm min}\leq \gamma \leq \gamma_{0}$} \\
            N_{0}\gamma _{\rm 0}^{-p} \left(\frac{\gamma}{\gamma_{0}}\right)^{-3}10^{-b\log^{2}(\gamma/\gamma_{0})}  &  \mbox{ $\gamma _{\rm 0}<\gamma\leq\gamma_{\rm max}$.}
           \end{array}
       \right.
\label{Ngamma}
\end{equation}

The emission region moves relativistically with a Lorentz factor $\Gamma=1/\sqrt{1-\beta^{2}}$ and a viewing angle $\theta$,
which forms the Doppler beaming factor $\delta=1/\left[\Gamma\left(1-\beta\cos\theta\right)\right]$. Frequency and
luminosity can be transformed from the jet frame to AGN frame as $\nu=\delta\nu'$ and $\nu L(\nu)=\delta^{4}\nu' L'(\nu')$,
respectively.
The prime refers to the values in the jet frame. Due to causality argument, the variability timescale set a constraint on the size of
emission region $R\lesssim\delta\Delta tc/(1+z)$. Because the epochs 2012 and 2013 are quiet
states, therefore for simplicity, variability timescale is set to be $\Delta t=10$ days during our SED modeling.
All jet parameters, except R are constrained through SED modeling.
The Klein-Nishina (KN) and SSA effects are also considered \citep[see][]{2017ApJ...842..129C}.

Recently, Kataoka \& Stawarz (2016) presented the NuSTAR observations of Mrk 421 and found X-ray spectra to be concave (hard
X-ray excess) in its low state. They explained this feature to be the power law extension of Fermi/LAT spectra.
However, \citet{2017ApJ...842..129C} explored the possibility whether the hard X-ray excess of Mrk 421 is produced from the
inverse Compton emission of the low energy end of the power law electron distribution. It is found that the hard X-ray excess
of Mrk 421 could be well fitted by the one zone model provided the minimum electron energy would be $\gamma_{min}\approx19$.
However, it is found that the predicted radio flux was significantly larger than the observed radio flux, even after
considering synchrotron self absorption effects \citep{2017ApJ...842..129C}. Due to the overprediction of the radio flux,
 it is implied that the observed hard X-ray excess of Mrk 421 could not be the low-energy tail of the Fermi/LAT $\gamma$-ray spectra.

Similarly, for PKS 2155--304, first we model the broadband SED (eyeball fit) for epochs 2012 and 2013 having hard X-ray 
excess using one zone model. We find that in order to fit these two concave X-ray spectra within the one zone model, minimum
electron energy required should be at least $\gamma_{min}\lesssim30$. The modeled curves are shown in Figure \ref{fig:sedonezone},
where we presented three SEDs with $\gamma_{\rm min}=10$, 30 and 100, respectively. The corresponding fitting parameters
for these $\gamma_{min}$ are provided in Table \ref{tab:jetparamters_onezone}.
It can be seen from the figure that due to the cutoff of SSC emissions at lower energy band, these emissions decreases
significantly at X-ray band with increasing $\gamma_{\rm min}$. Even up to $\gamma_{\rm min}=100$, the predicted
 radio flux is significantly larger than the observed one. Similar to Mrk 421 \citep{2017ApJ...842..129C}, it
implies that the electron power-law distribution cannot extend down to such low energies (i.e., $\gamma_{\rm min}\lesssim30$).
Hence, the observed X-ray concave feature of PKS 2155--304 cannot be produced by the low-energy part of the same electron
population that produced the {\it Fermi}/LAT $\gamma$-rays. It can be noted that the SED of PKS 2155--304 during 2013 was
also modeled by \citet{2016ApJ...831..142M} using the one zone model and the derived minimum electron energy is
$\gamma_{min}\sim1$. But, it can noted that they are lacking the radio data in their SED modeling \citep[see Figure 4
in][]{2016ApJ...831..142M}.

The set of variability time scale may effect our results. Therefore, we set $T_{var}$ = 1 day and model the SED again, which 
leads to smaller R and larger B values. In this case, the Doppler factor reaches very large values ($\delta$ = 47 for epoch 2012 
and $\delta$ = 58 for epoch 2013). The modeling curves are presented in Figure 6 as dash-dotted black lines. It can be seen that 
even in this extreme case, the predicted radio flux is larger than the observed one.

\begin{table*}
\textwidth=6.0in
\centering
\caption{The jet parameters for one zone SED modeling for PKS 2155-304. The model curves for different $\gamma_{min}$
 are shown in Figure \ref{fig:sedonezone}. 2012-1 and 2013-1 are SED parameters modeled with $\Delta t=10$ days and 2012-2 
and 2013-2 are SED parameters modeled with $\Delta t=1$ day (see text for details).} 
\begin{tabular}{@{}lcccccccccccc@{}}
\hline
\hline
     & R (10$^{17}$cm) & $\delta$ & $B$ (10$^{-2}$ Gs) &  $N_0$ & $\gamma_0$ & $\gamma_{\rm max}$ & $p$ & $b$\\
\hline
2012-1 & 6.84 & 23.8 & 0.76 & 944  & 137256 & 100$\gamma_{0}$& 2.5 & 1.0 \\
2012-2 & 1.36 & 47.1 & 1.05 & 5143  & 82704 & 100$\gamma_{0}$& 2.5 & 1.0 \\
2013-1 & 8.19 & 28.3 & 0.96 & 150  & 80822  & 100$\gamma_{0}$& 2.4 & 0.75 \\
2013-2 & 1.68 & 58.2 & 1.20 & 831  & 50412  & 100$\gamma_{0}$& 2.4 & 0.75 \\
\hline
\end{tabular}  \\
\label{tab:jetparamters_onezone}
R: Size of the emission region \\
$\delta$: Doppler factor \\
$B$: Magnetic field strength\\
$N_0$: Normalized electron number density \\
$\gamma_0$: broken electron energy   \\
$\gamma_{\rm max}$: Maximum electron energy  \\
$p$: Spectral index \\
b: Curvature \\
\end{table*}

\subsubsection{The spine/layer jet}
\label{sec:sed_spine_layer_model}

As detected by H.E.S.S., PKS 2155-304 has shown very rapid variability at TeV band and the minimum variability timescale can be
as small as $t_{\rm var}=3-5$ minutes \citep[][Another HSP BL Lac Mrk 501 also has minutes variability at TeV band as revealed by
MAGIC Telescope, \citet{2007ApJ...669..862A}]{2007ApJ...664L..71A}. These fast variability of blazars indicate substantial
sub-structure in their jets, which may be due to turbulence or as a result of magnetic reconnection \citep{2008MNRAS.386L..28G,
2009MNRAS.395L..29G, 2012A&A...545A.125R}. Rapid variability also implies significant beaming effect of TeV HSP BL Lacs including
PKS 2155-304. However, it is very interesting that the high resolution VLBA observations of PKS 2155-304 did not find superluminal
motion of the central jet \citep{2008ApJ...678...64P}. One competitive possibility is that the jet has an inhomogeneous structure
transverse to the jet axis, consisting of a fast `spine" that dominates the high-energy emission and a slower `layer" that
dominates at lower frequencies \citep{2006ApJ...640..185H, 2005A&A...432..401G, 2004ApJ...613..752G}, which is consistent with the
above sub-structure of the jet. The spine/layer jet model was initially studied by \citet{1989MNRAS.237..411S}.
Irrespective of the origin of such transverse structures, their observational signatures are visible in
the VLBI images of jets that are resolved in the transverse direction, either in the form of limb brightening or limb darkening (according
to the region dominated by the radio emission). Observational signatures of limb brightening have been claimed in the VLBI
images of the TeV sources such as Mrk 501 \citep{2004ApJ...600..127G}, Mrk 421 \citep{2006ApJ...646..801G} and M87
\citep{2007ApJ...668L..27K}. As suggested by \citet{2008ApJ...678...64P}, VLBA images of PKS 2155-304 is not detected with sufficient
resolution or dynamic range to discern structure transverse to the jet axis. Such observations are limited to the
brighter sources (or to much more sensitive images of such sources).

Here, we model the broadband emission of PKS 2155-304 having concave X-ray spectra with the spine/layer model. We present very brief
introduction of the model, please see \citet{2017ApJ...842..129C} for its detailed description.
The spine/layer model used here is same as described in \citet{2005A&A...432..401G}. For geometry, see Figure 1 of
\citet{2005A&A...432..401G}. The layer is assumed to be a hollow cylinder with external radius $R_{2}$, internal radius $R$ and
width $\Delta R_{\rm l}^{''}$ in the comoving frame of the layer\footnote{Double prime refers to values in the frame of the layer,
and prime refers to values in the frame of the spine.}. For the cylindrical spine, the radius is $R$ and the width is
$\Delta R_{\rm s}^{'}$. The spine and layer move with velocities $c\beta_{\rm s}$ and $c\beta_{\rm l}$, respectively, with
$\Gamma_{\rm s}$ and $\Gamma_{\rm l}$ are the corresponding Lorentz factors. The relative velocity between the spine and the layer
is then $\Gamma_{\rm rel}=\Gamma_{\rm s}\Gamma_{\rm l}(1-\beta_{\rm s}\beta_{\rm l})$. Following \citet{2005A&A...432..401G}, the
radiation energy densities are considered as follows,

\begin{itemize}
  \item In the comoving frame of the layer, the radiation energy density $U_{\rm l}^{''}=L_{\rm l}^{''}/\left[\pi \left(R_{2}^{2}-R^{2}\right)c\right]$ \citep[slightly different from that of][to make sure that the radiation energy density is in the same format
as that in the spine]{2005A&A...432..401G}. In the frame of the spine, this radiation energy density will be boosted
to $U_{\rm l}^{'}=\Gamma_{\rm rel}^{2}U_{\rm l}^{''}$.
  \item In the comoving frame of the spine, the radiation energy within the spine is assumed to be $U_{\rm s}^{'}=L_{\rm s}^{'}/\left(\pi R^{2}c\right)$. The radiation energy density observed in the frame of the layer will be boosted by
$\Gamma_{\rm rel}^{2}$ but also diluted (since the layer is larger than the spine) by a factor
$\Delta R_{\rm s}^{''}/\Delta R_{\rm l}^{''}=\left(\Delta R_{\rm s}^{'}/\Gamma_{\rm rel}\right)/\Delta R_{\rm l}^{''}$.
\end{itemize}

The electron energy distributions in the spine and layer are all assumed to be broken power law plus log-parabolic model, 
as expressed in Equation \ref{Ngamma}, but with different parameters. In our modeling, we always assume
 $\Delta R_{\rm l}^{''}=30\Delta R_{\rm s}^{'}$, as given in \citet{2005A&A...432..401G}. Then, we model the SEDs of 
epochs 2012 and 2013 by employing the
spine/layer model. Figure \ref{fig:sedspinelayer} presents our SED modeling results, and the corresponding parameters are
listed in Table \ref{tab:jetparamters_spinelayer}. Geometry parameters for epoch 2012 and 2013 are $R=1.38\times10^{18}$ cm,
$R_{2}=1.65\times10^{18}$ cm, $\Delta R_{\rm s}^{'}=2.29\times10^{17}$ cm and $R=1.62\times10^{18}$ cm, $R_{2}=1.94\times10^{18}$ cm,
 $\Delta R_{\rm s}^{'}=2.69\times10^{17}$ cm, respectively. For both SEDs, red and blue lines represents the spine
and layer emissions, respectively; dot-dashed lines represents synchrotron emissions, dotted lines represents the SSC
emissions, and dashed lines represents IC emissions of seed photons originated externally from the layer/spine. It can
been seen that, similar to one-zone modeling, the synchrotron $+$ SSC emissions of the spine can well re-produce almost whole SED,
except for the hard X-ray excess. The rise part of X-ray concave spectra can be successfully represented by the process of
seed photons (produced from the layer) being IC scattered by the non-thermal electrons within the spine.

\begin{table*}
{\caption{The jet parameters for spine/Layer modeling for PKS 2155-304. The model curves are shown in Figure 7.}}
\textwidth=6.0in
\scriptsize
\setlength{\tabcolsep}{0.035in}
\noindent
\centering
\begin{tabular}{@{}lcccccccccccc@{}}
\hline
\hline
     & $\theta$ ($^{\circ}$) &  $\Gamma$&  $\delta$ & $B$ (10$^{-2}$ Gs) &  $N_0$ & $\gamma_0$ & $\gamma_{\rm min}$& $\gamma_{\rm max}$ & $p$ & $b$\\
\hline
Spine(2012) & 1 & 12.5 & 23.8 & 0.75  & 2554 & 154004 & 3000 & 1000$\gamma_{0}$& 2.6 & 1.0 \\
Layer(2012) & 1 & 3.26 & 6.34 & 0.065 & 232  & 2646   & 2    & 1000$\gamma_{0}$& 2.0 & 1.0 \\
Spine(2013) & 1 & 14.9 & 28.0 & 0.95  & 1070 & 143724 & 3000 & 1000$\gamma_{0}$& 2.6 & 1.0 \\
Layer(2013) & 1 & 3.49 & 6.80 & 0.076 & 133  & 2498   & 2    & 1000$\gamma_{0}$& 2.0 & 1.0 \\
\hline
\end{tabular} \\
\label{tab:jetparamters_spinelayer}
$\theta$ ($^{\circ}$): Jet veiwing angle  \\
$\Gamma$: Lorentz factor    \\
$\delta$: Doppler factor  \\
$B$: Magnetic field strength          \\
$N_0$: Normalized electron number density  \\
$\gamma_0$: Broken electron energy  \\
$\gamma_{\rm min}$ and $\gamma_{\rm max}$: Minimumm and maximum electron energy            \\
$b$:  curvature   \\
\end{table*}

\section {Discussion \& Conclusions}

We presented the results of spectral analysis of PKS 2155$-$304 using the XMM$-$Newton observations during the period
2009--2014.

 We analysed spectra of all of the pointings of PKS 2155--304 and fit each spectra in the range 0.6--10 KeV using single power-law
 and log-parabolic model. We found that log-parabolic model better describes the most observations.
From figure 4, it is clear that in most occasions there are residuals in the upward direction indicating flattening of the
spectra above 6--7 KeV, but the spectral fits do not show negative curvature except in 2012. This pointing has the lowest
flux among all of the observations studied here. Such type of spectral flattening of PKS 2155--304 is also found in previous
studies by Zhang (2008) and Foschini et al. (2008) and is interpreted as a mixture of the high energy tail of the synchrotron
component which is contaminated by low-energy end of the IC component.  Using NuSTAR Satellite, Kataoka \& Stawarz (2016)
and Madejski et al. (2016) found the X-ray concave spectra of well known HBL Mrk 421 and PKS 2155--304 respectively.
They explained this feature to be the power law extension of the Fermi/LAT spectra.
Since, the observations of Madejski et al. (2016) in 2013 are simultaneous with our observations of XMM--Newton, we included their data
in our SED fitting.

In this way, we have significant X-ray concave spectra during epochs 2012 and 2013. Out of many possible explanation for the observed
X-ray excess, one is that this excess is actually the high-energy tail of synchrotron X-ray emissions.
But, it requires a spectral pile-up in electron distribution at the highest
energies. This seems unlikely to work. The main reasons are as follows. (1) This high-energy pile-up bump
requires the acceleration timescale to be equal to the radiative-loss timescale at the limit for the perfect confinement
of electrons within the emission zone \citep{2008ApJ...681.1725S}. However, at lower electron energies this scenario predicts
 a flat power tail, which is inconsistent with the X-ray observations of PKS 2155--304. (2) This high-energy tail can
also be achieved by the electron high-energy pile-up due to the reduction of the IC cross-section in the KN regime
\citep{2005MNRAS.363..954M}. However, it requires IC cooling dominating over synchrotron cooling. It is not consistent
with the fact that PKS 2155-304 is almost totally dominated by synchrotron emission. Alternatively, it seems to be the low-energy
tail of the {\it Fermi}/LAT $\gamma$-ray spectra. As discussed in Section \ref{sec:sed_one_zone_model}, this possibility requires the
minimal electron energy to be $\gamma_{\rm min}\lesssim30$. Such a lower electron energy predicts a very strong radio
emission (after considering SSA effect), which is larger than the observed radio flux. Therefore, the hard X-ray excess
cannot be the low-energy tail of the {\it Fermi}/LAT $\gamma$-ray spectra.

In this paper, we explore the possibility whether the origin of this concave X-ray feature is related to the spine/layer jet
structure: a fast spine surrounded by a slower layer. The main reason for the choice of spine/layer model is due to the
fact that PKS 2155-304 shows rapid variability and very slow radio proper motion (see Subsection \ref{sec:sed_spine_layer_model}),
which indicates that its jet may structured. Also, this model is able to produce IC SEDs that are significantly different from
those produced by the one-zone model \citep[][]{2005A&A...432..401G, 2016MNRAS.457.1352S, 2017ApJ...842..129C}.
Spine/layer model can explain some inconsistencies between observations and the AGN standard unified model. According to the
standard unified model, blazars and radio galaxies appear different due to difference in the viewing angles
\citep{1995PASP..107..803U}. Since, Doppler beaming factor depends significantly on the viewing angle. Hence, one can obtain
 number distribution (number ratio between radio galaxies and blazars) as a function of viewing angle. It is found that within
the one zone model, this number ratio is inconsistent with observations \citep{2000A&A...358..104C}.
 This inconsistency can be explained using the spine/layer model e.g., a faster-moving spine surrounded
by a slower-moving layer \citep{2000A&A...358..104C}. In this model, sources having jet with smaller viewing angles with
respect to our line of sight will be dominated by the faster spine (i.e. blazars), while the layer component will contribute more or
even dominate the emissions in the sources having larger viewing angle (i.e. for radio galaxies). Hence, spine/layer model predicts
larger number ratio between blazars and radio galaxies as compared to the one-zone model \citep{2000A&A...358..104C}. Also,
the relative motion between the spine and the layer will amplify the photon energy density produced from the spine
(layer) in the frame of the layer (spine) by $\sim\Gamma_{\rm rel}^{2}$ (see Subsection \ref{sec:sed_spine_layer_model}).
and therefore, the IC emissions will be enhanced in this model as compared to one-zone model \citep{2005A&A...432..401G}.

We find that the spine emissions provide good modeling of the Multi-wavelength SED of PKS 2155--304. However, similar to
one zone modeling, the hard X-ray spectra is not well fitted by spine emissions.  Similar to Mrk 421 \citep{2017ApJ...842..129C},
the rising part of concave X-ray spectra can be well
represented by the synchrotron photons from the layer being IC scattered by the spine electrons (see Figure \ref{fig:sedspinelayer}).

Until now, about 49 HSP BL Lacs\footnote{http://tevcat.uchicago.edu/} have been detected with VHE emissions. Also, orphan flare
is an interesting and less known phenomenon in blazar observations. The spine/layer model has been successfully used to explain
the VHE $\gamma$-ray emission of blazars \citep[see, e.g., ][]{2005A&A...432..401G, 2008MNRAS.385L..98T, 2016MNRAS.457.1352S}.
In order to explain the orphan flare in the $\gamma$-ray band, it is proposed that, when a faster-moving spine moves across a
slower-moving layer/ring, the sudden increase of the energy density of external seed photons
(produced from the layer) will be IC scattered by the spine electrons \citep{2015ApJ...804..111M, 2016arXiv161109953M}.
The origin of the spine/layer structure is not yet clear. However, in some numerical simulations of
\citet[e.g.,][]{2007ApJ...662..835M, 2010A&A...521A..67M} it is found that it could arise directly from a jet launching process
in which the external layer is ejected from the accretion disk while the central spine is fueled from the black hole ergosphere.
Another possibility is accumulation of the inflating toroidal magnetic field inside the accretion disk due to the magnetized accretion
flow. It can produce a magnetic tower jet which presents a central helical magnetic field ``spine" surrounded by a reversed magnetic field
``sheath," which could be a prototype of a spine/layer structure \citep[see, e.g.,][]{2004ApJ...605..307K, 2006ApJ...643...92L}.

The Multi-wavelength SED of most of the HSP BL Lacs are  well fitted by the one-zone model \citep[e.e.,][]{2009A&A...504..821P,
2012ApJ...752..157Z, 2014Natur.515..376G, 2014MNRAS.439.2933Y, 2017MNRAS.464..599D}. But, due to the limited sensitivity
of the hard X-ray instruments, the rising/concave part of the X-ray band is not included in the fitting.
If more sources like Mrk 421 and PKS 2155--304 are observed with hard X-ray spectra, it might be possible that their SED modelings
underestimate their hard X-ray emissions and require spine/layer model to better constrain them. For our SED modeling of PKS 2155-304,
it should be noted that the model used is just a simple toy model, i.e., using two distinct regions (spine and layer) instead of
continuing the distribution from jet axis to edge, although our modeled parameters are among the typical values of blazars
\citep[e.g.,][]{2010MNRAS.402..497G, 2014Natur.515..376G, 2012ApJ...752..157Z}.

\begin{acknowledgements}
We would like to thanks the anonymous reviewer for the constructive comments which helped us to improve the
manuscript scientifically.
 We acknowledge Prof. A.E. Volvach for kindly providing the radio data. HG acknowledges Prof. F. Massaro for providing 
useful suggestions.  HG acknowledges Drs A. C. Gupta and J. C. Pandey for useful discussions. HG thanks Jithesh V., Shruti Tripathi, 
Nilkanth Vagshette and Mainpal Rajan for various helps and discussions
on spectral analysis. This research is based on observations obtained with XMM$-$Newton, an ESA science mission with instruments
and contributions directly funded by ESA member states and NASA.
H.G. is sponsored by the Chinese Academy of Sciences (CAS) Visiting Fellowship for Researchers from Developing Countries,
 CAS Presidents International Fellowship Initiative (grant No. 2014FFJB0005), supported by the NSFC Research
Fund for International Young Scientists (grant No. 11450110398) and supported by a Special Financial Grant from the China
Postdoctoral Science Foundation (grant No. 2016T90393).
L. Chen is supported by the NSFC (grants 11233006 and U1431123) and the CAS grant (QYZDJSSW-SYS023).
M. F. Gu is supported by the NSFC (grants 11473054 and U1531245).
\end{acknowledgements}

{

\begin{figure*}
\centering
\includegraphics[width=0.26\textwidth]{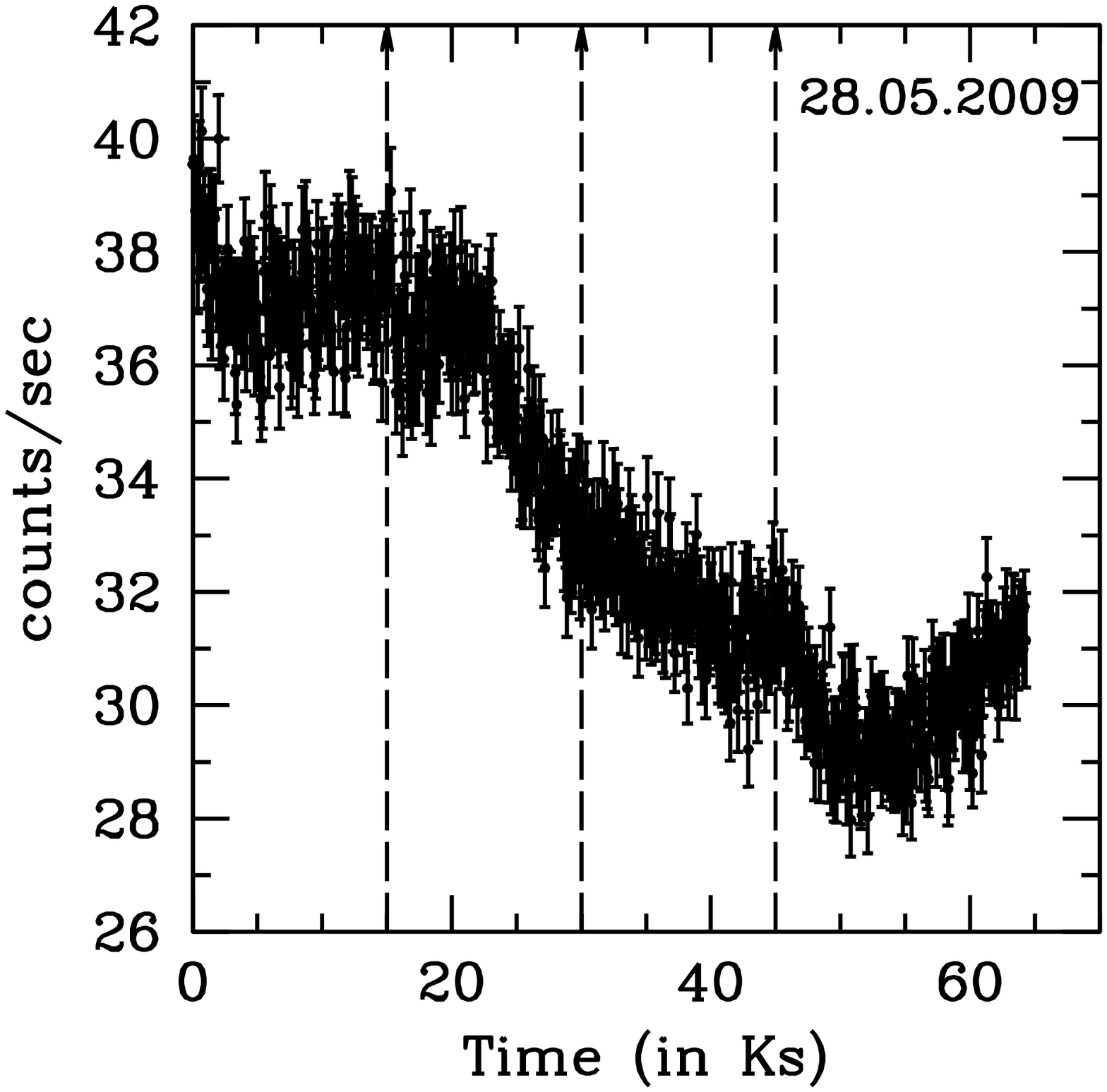}
\includegraphics[width=0.26\textwidth]{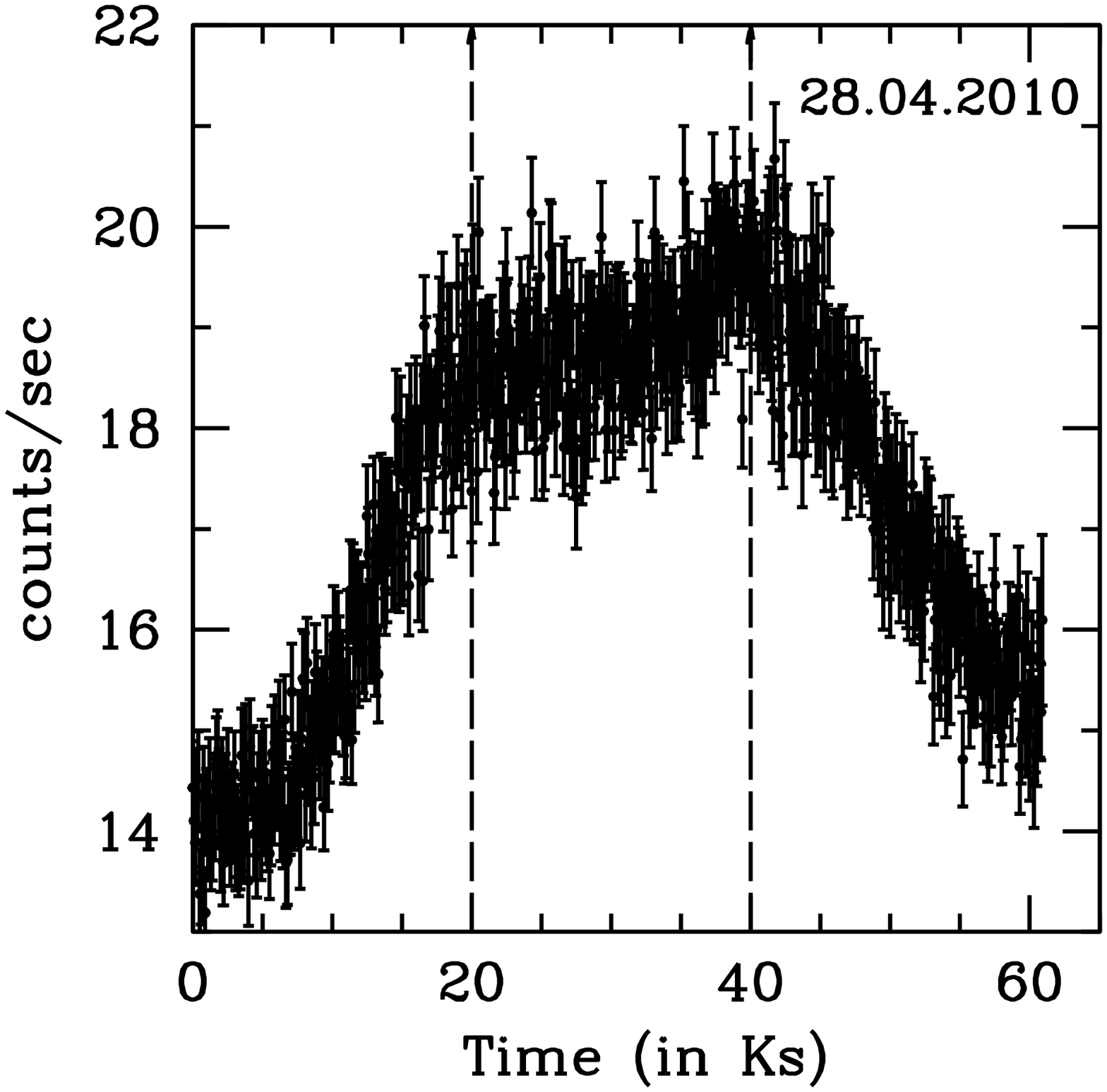}
\includegraphics[width=0.26\textwidth]{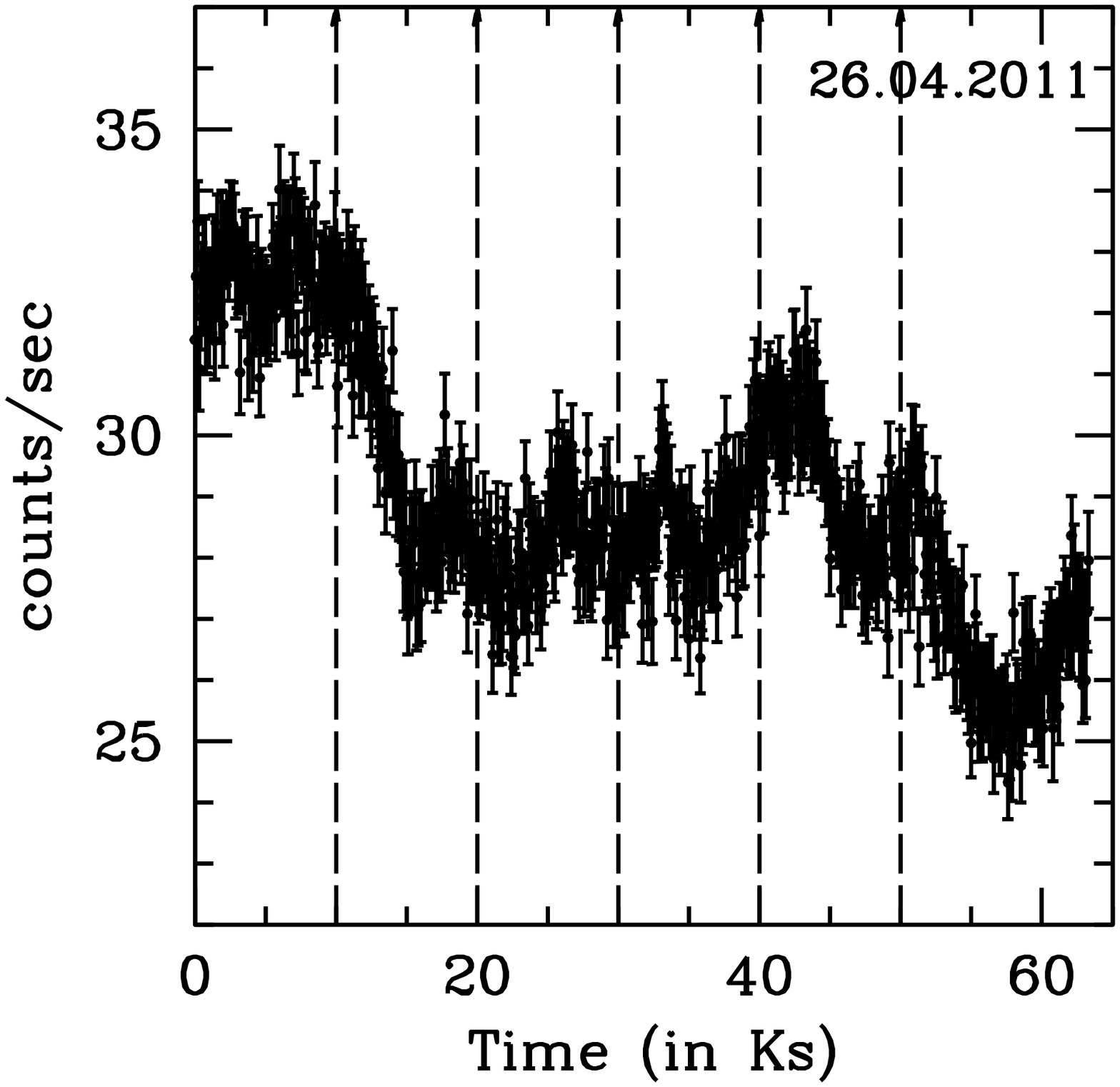}
\includegraphics[width=0.26\textwidth]{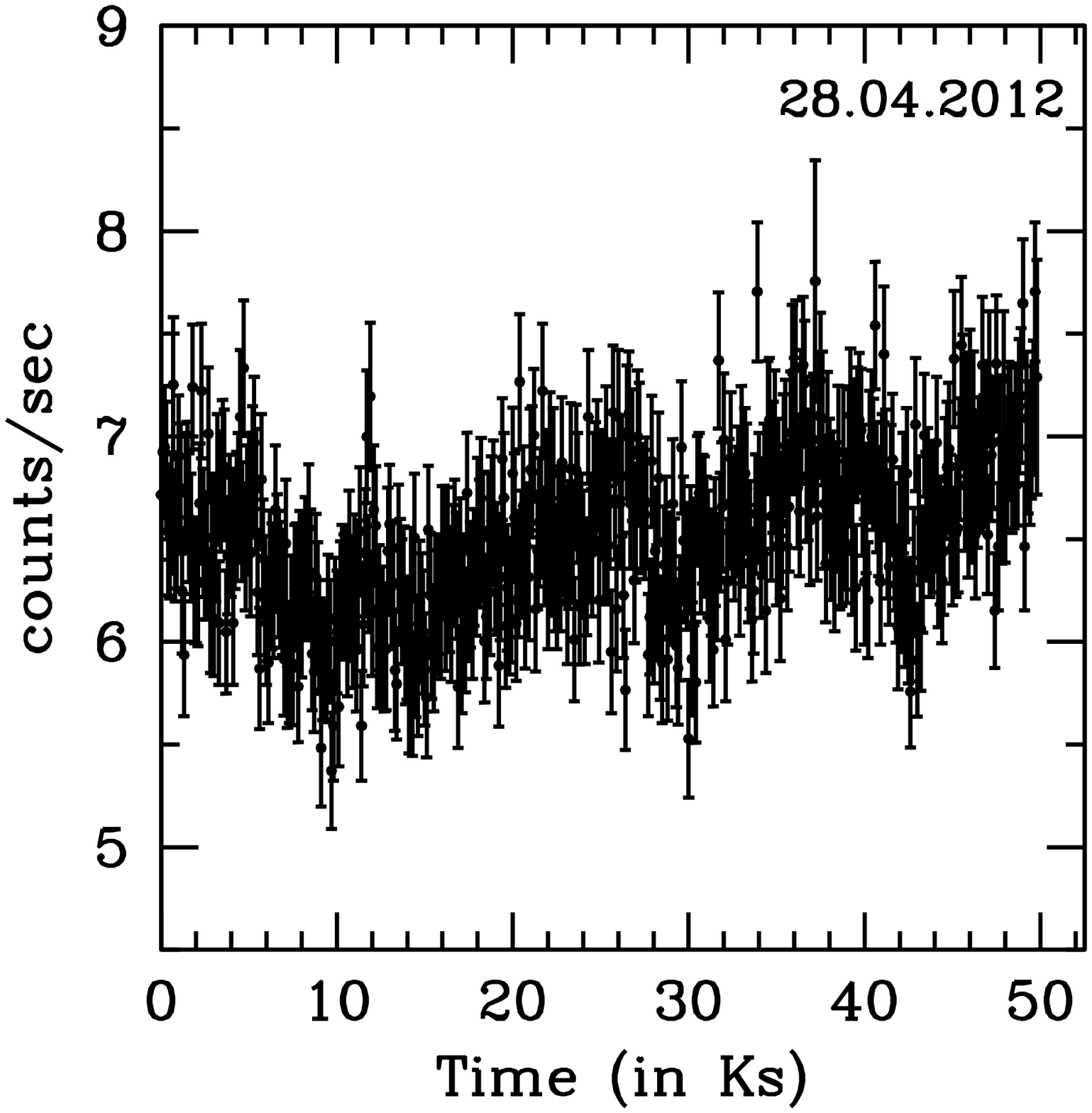}
\includegraphics[width=0.26\textwidth]{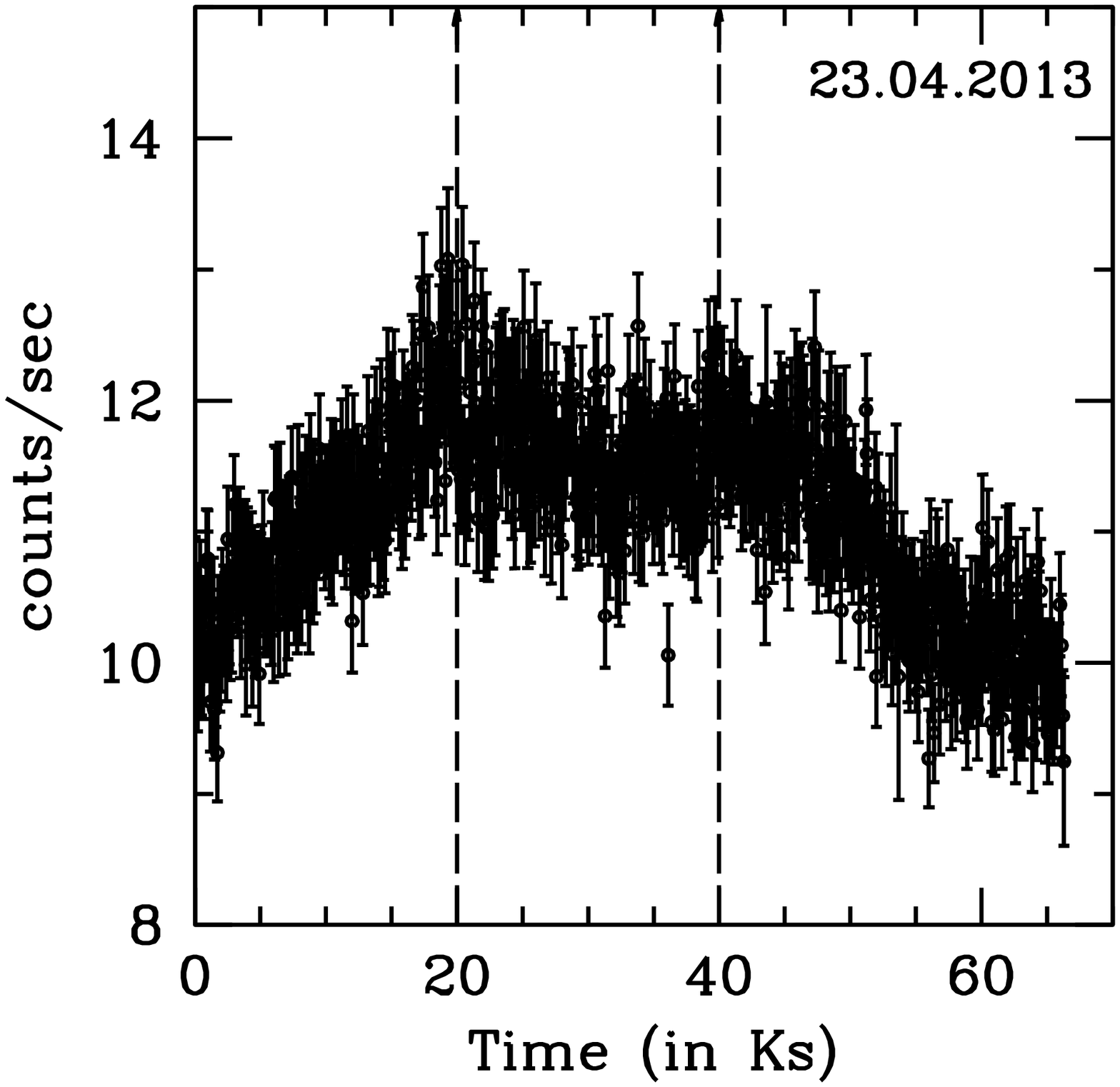}
\includegraphics[width=0.26\textwidth]{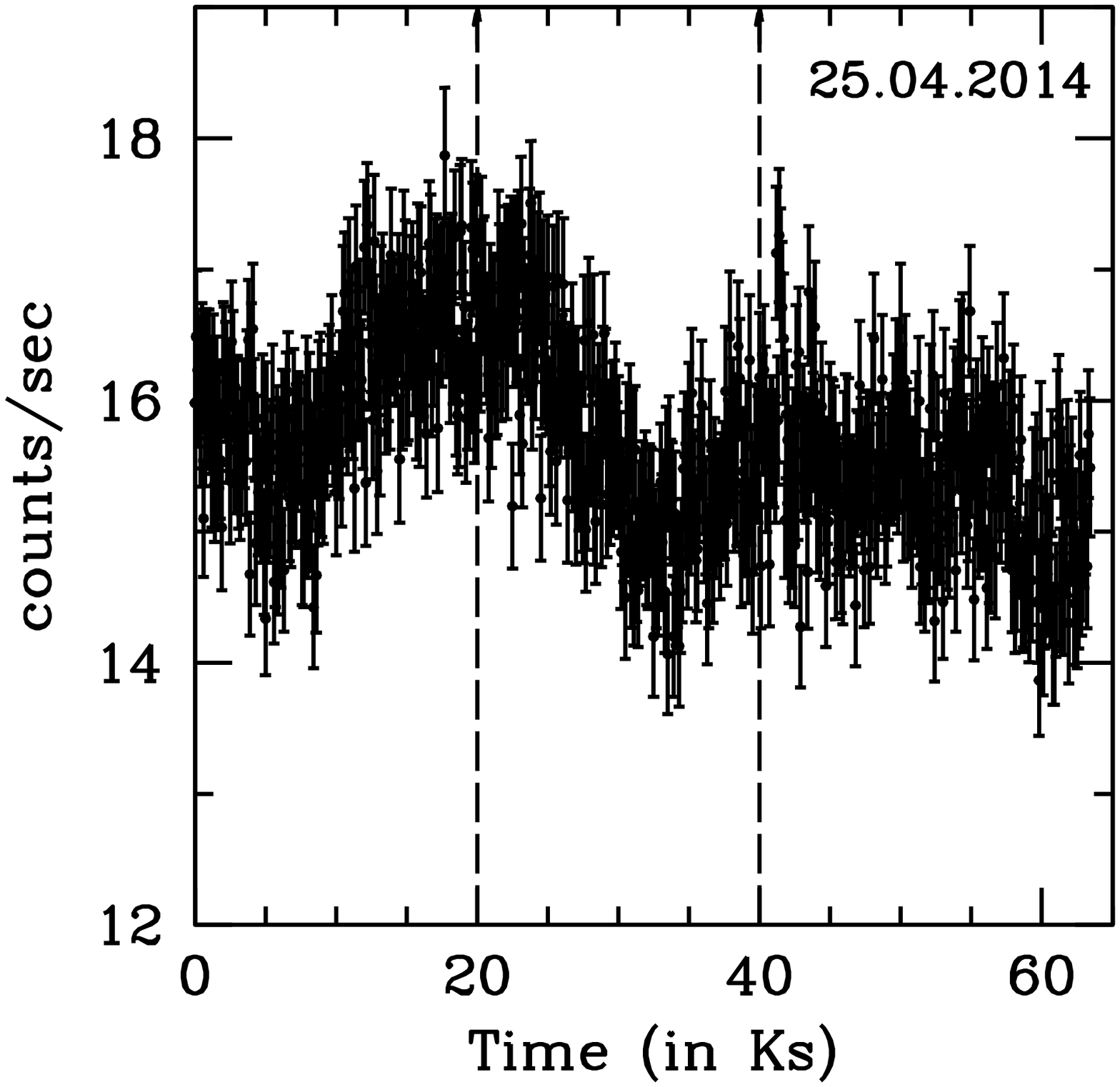}
\caption{The light curves of the blazar PKS 2155--304 during the period May 2009--April 2014
in 0.6--10 KeV. }
\end{figure*}

\begin{figure}
\centering
\includegraphics[width=0.8\textwidth]{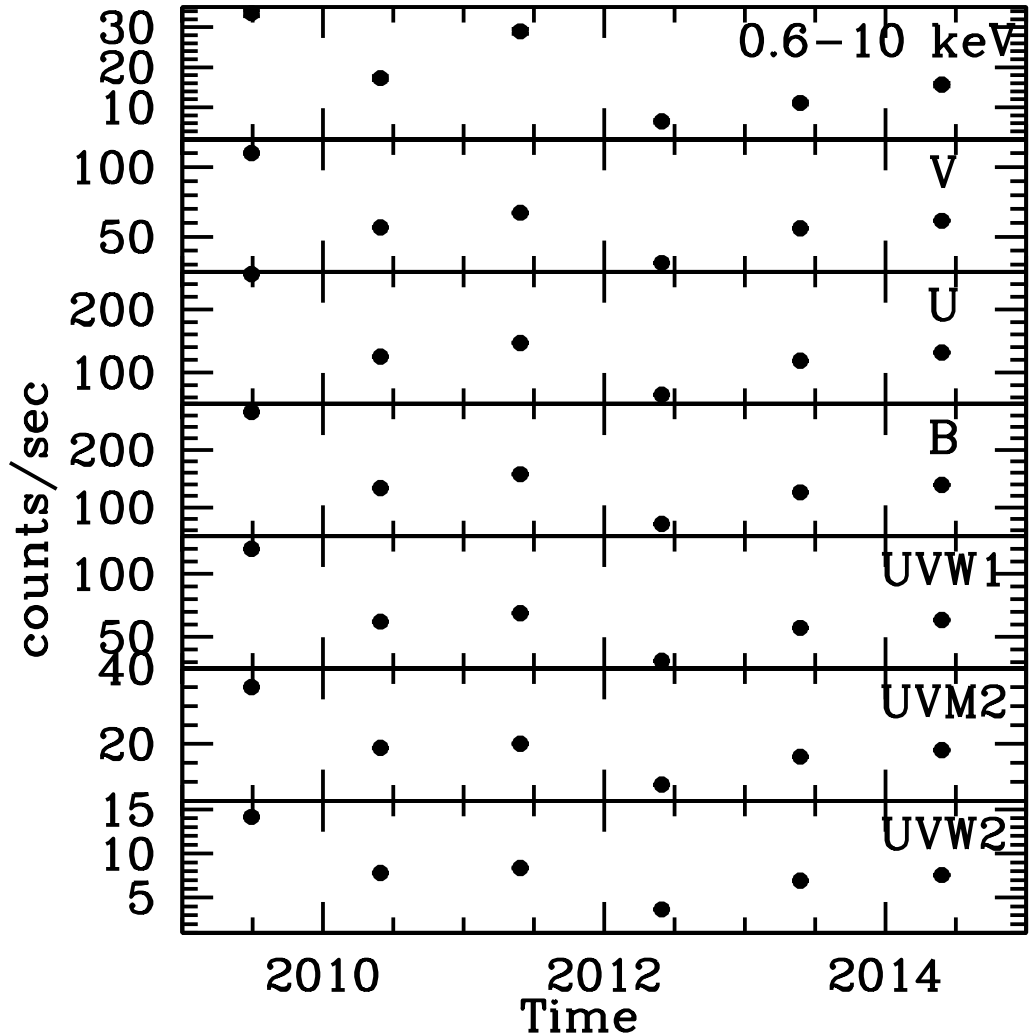}
\caption{XMM--Newton light curve of PKS 2155--304 during 2009--2014.}
\end{figure}

\begin{figure}
\centering
\includegraphics[width=0.8\textwidth]{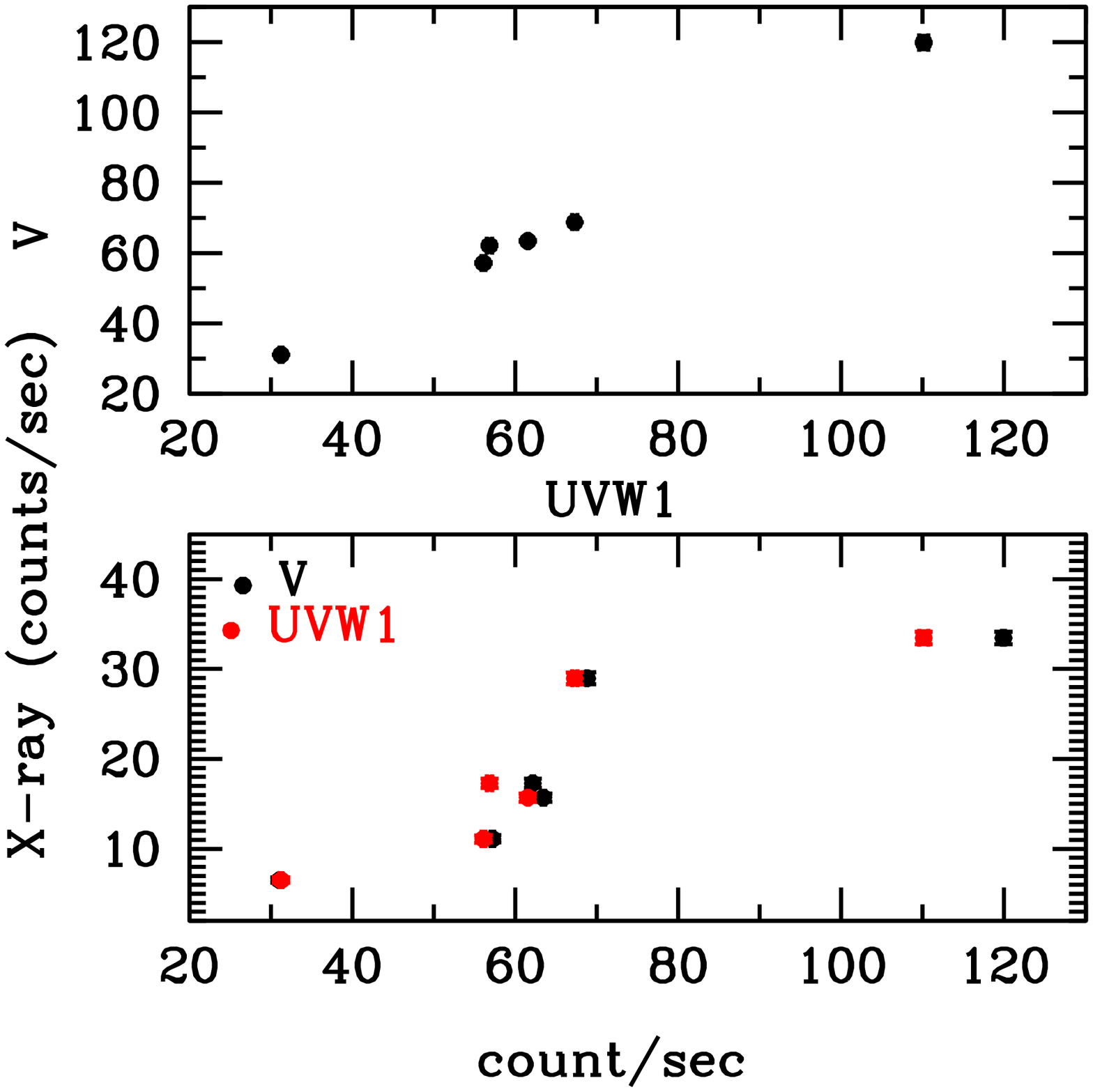}
\caption{Optical versus UV (upper panel) and X-ray versus optical/UV (lower panel).}
\end{figure}

\begin{figure*}
\centering
\includegraphics[width=0.45\textwidth]{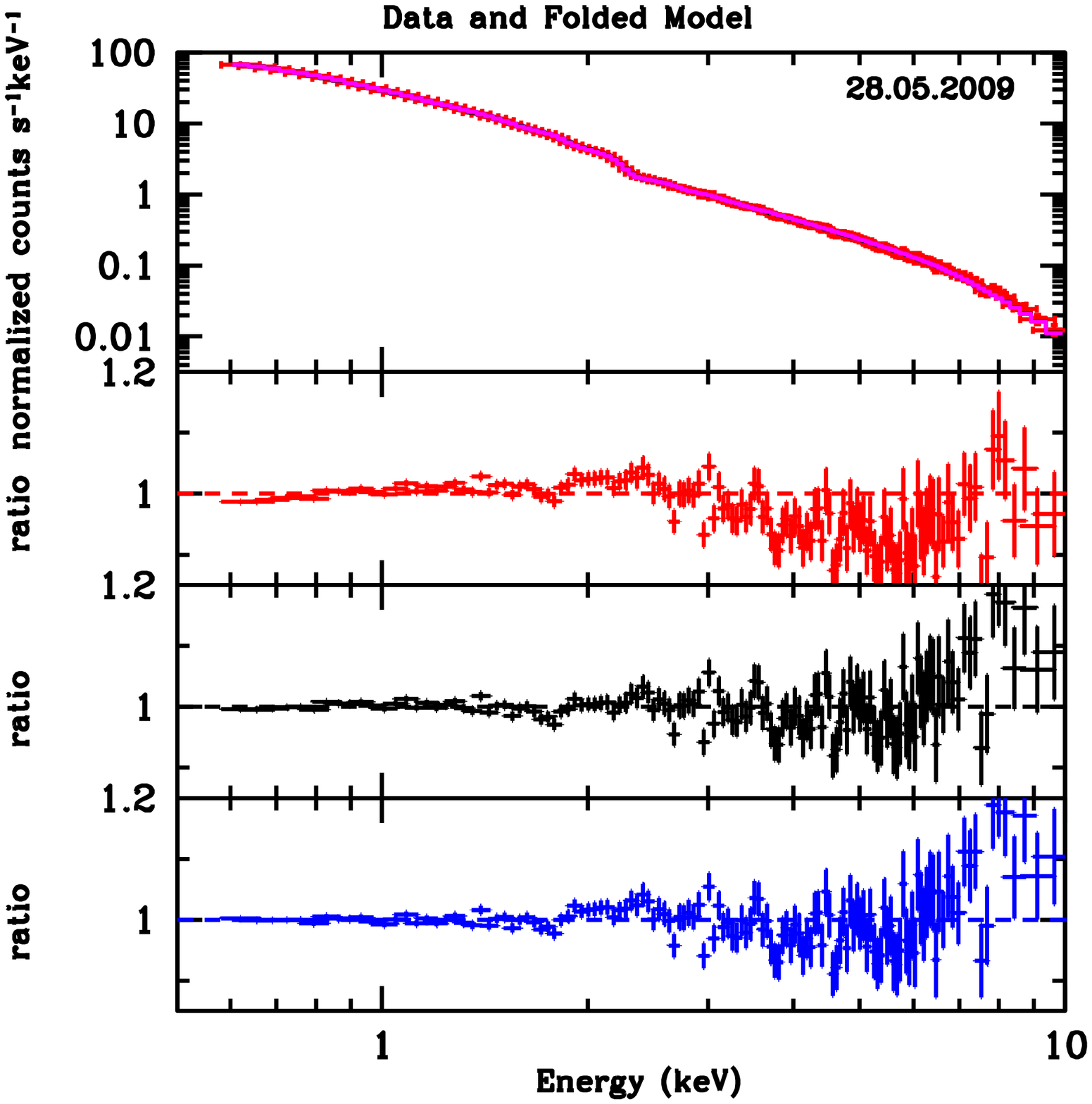}
\includegraphics[width=0.45\textwidth]{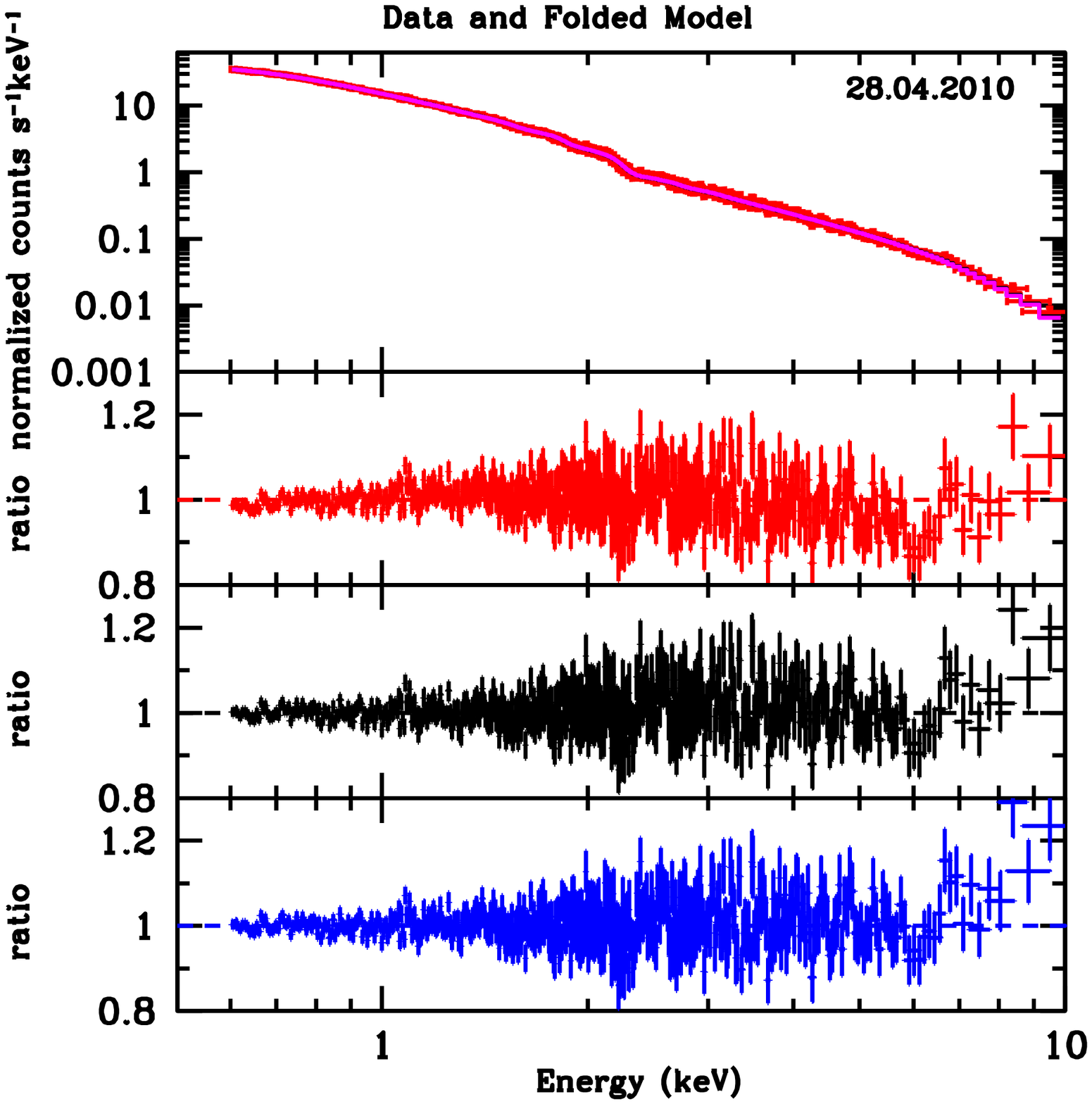}
\includegraphics[width=0.45\textwidth]{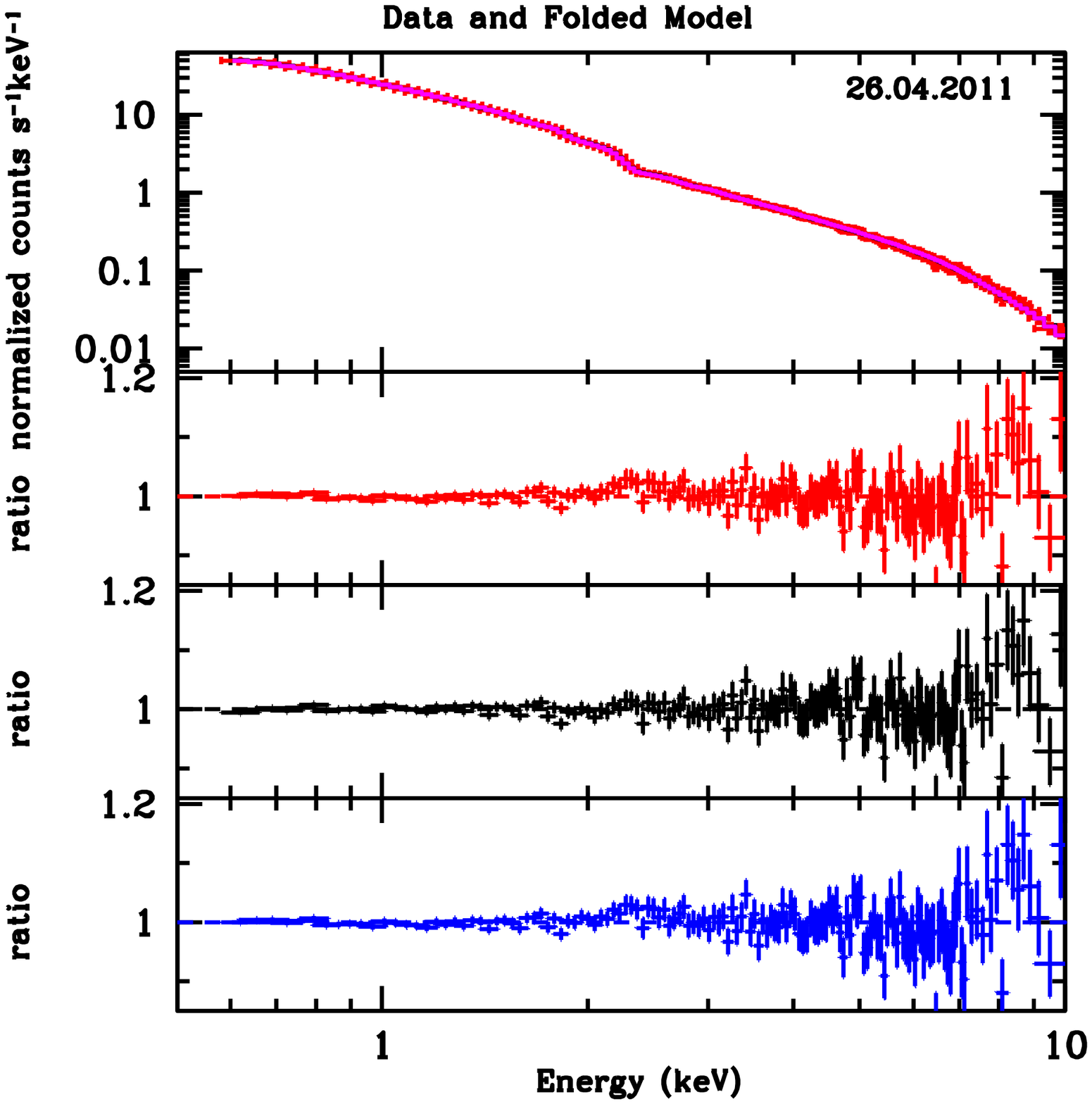}
\includegraphics[width=0.45\textwidth]{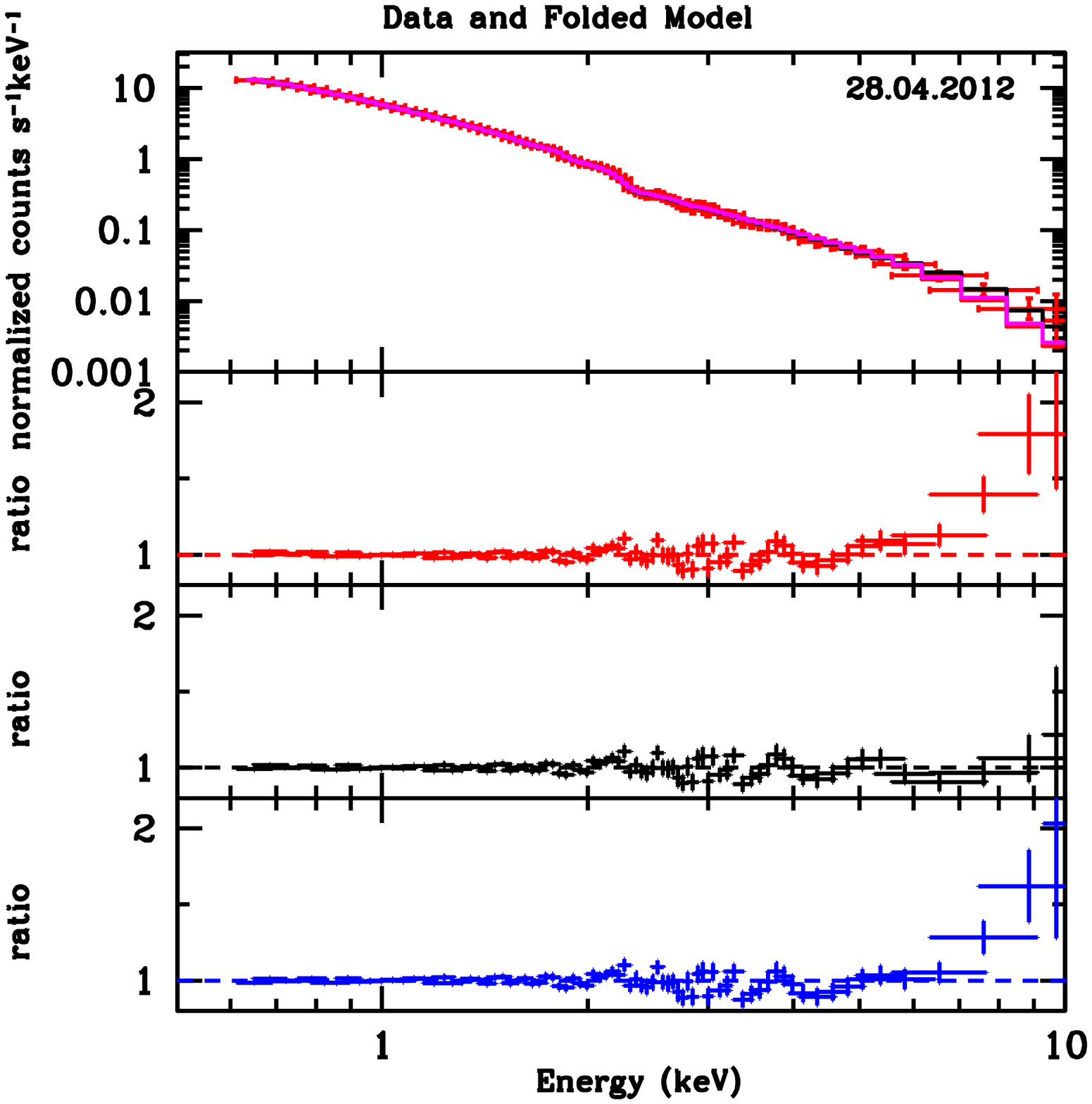}
\includegraphics[width=0.45\textwidth]{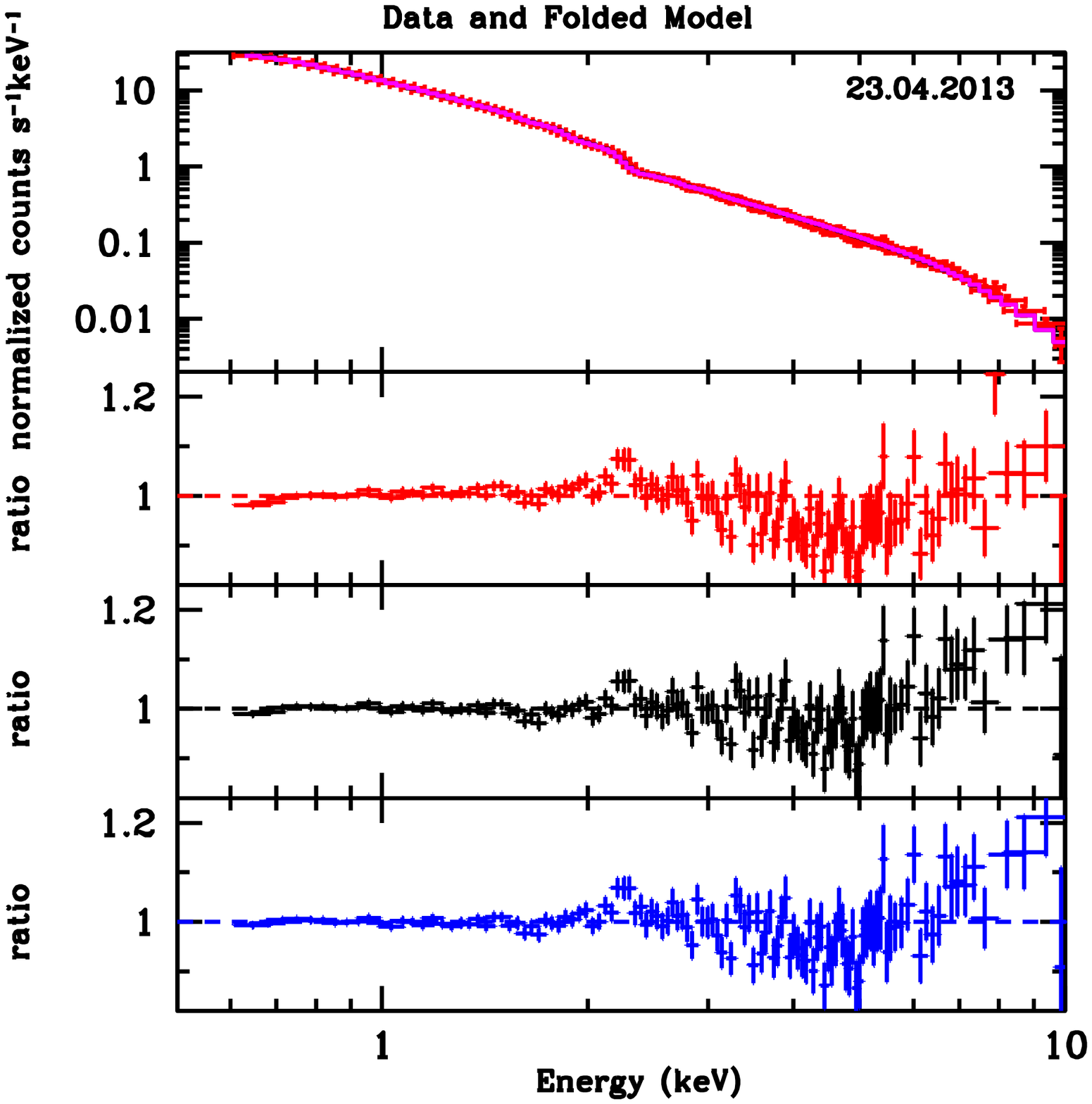}
\includegraphics[width=0.45\textwidth]{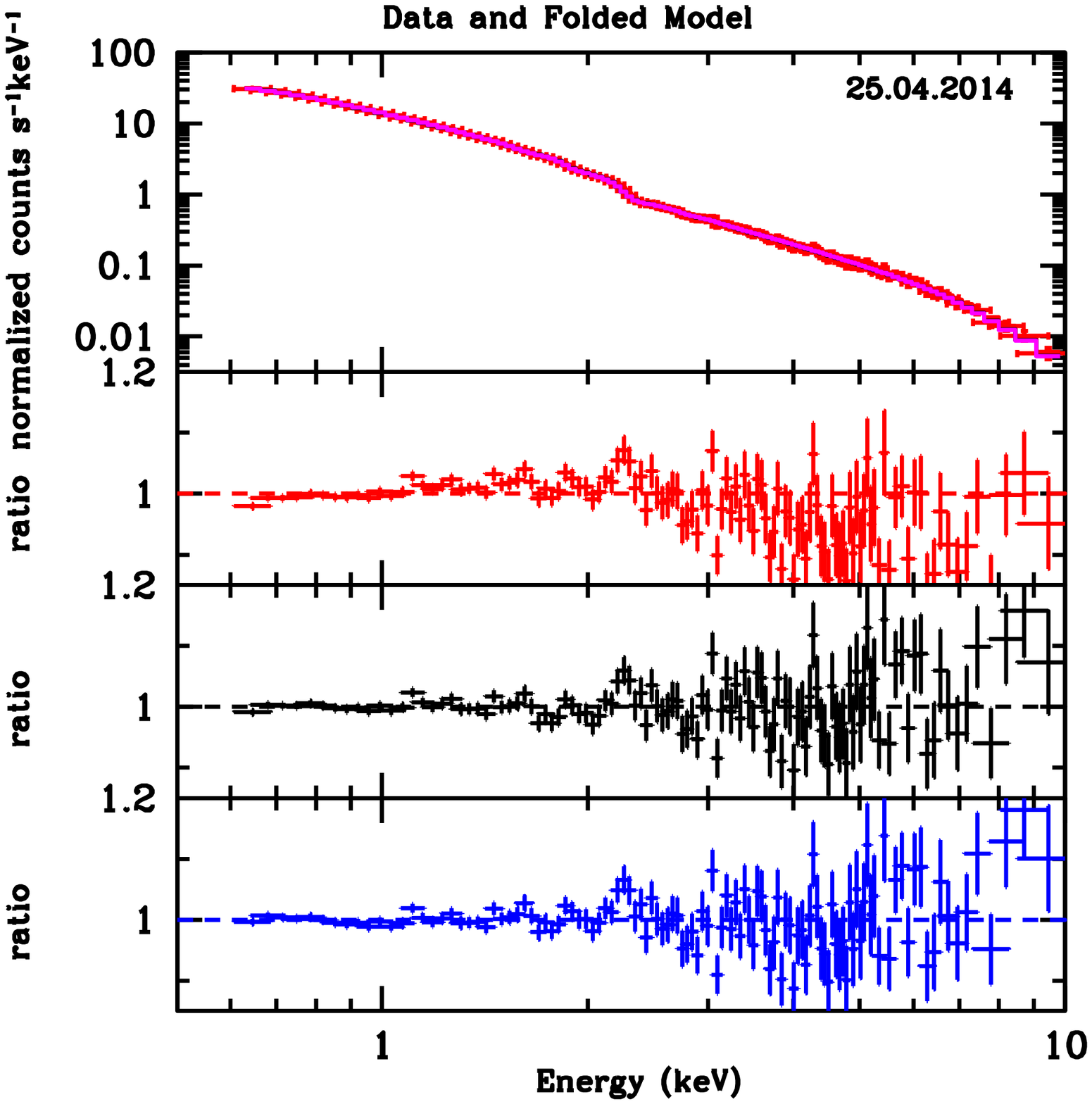}
\caption{The spectral fitting of six observations of PKS 2155--304 in 0.6--10 KeV. Each spectra is fitted
using the power-law, broken power law and log parabolic model and the data-to-model ratio is shown in the
three subpanels for each spectra in red, black and blue colors, respectively.}
\end{figure*}

\begin{figure}
\begin{center}
\includegraphics[width=0.85\textwidth]{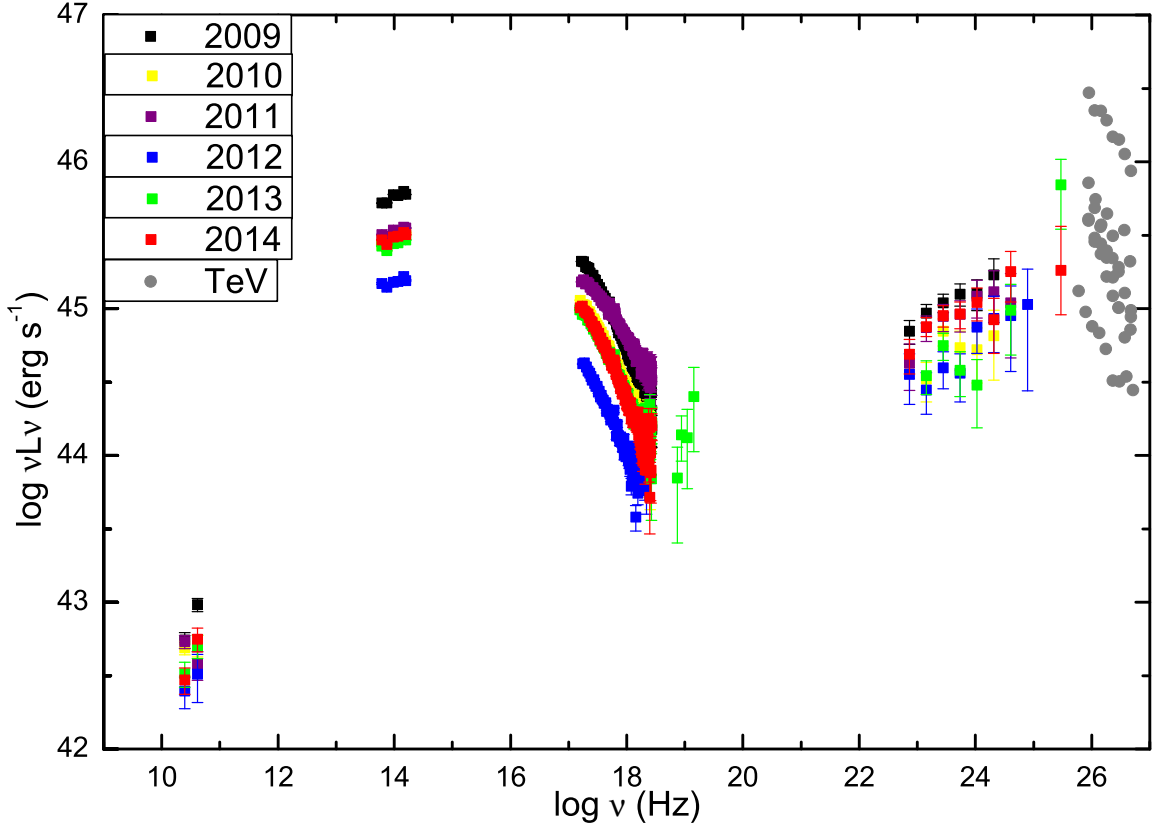}
\end{center}
\caption{Simultaneous broadband SEDs of PKS 2155-304 from epochs 2009-2014. The grey data are the non-simultaneous
TeV data from \citet{2012A&A...539A.149H}, corrected for EBL absorption \citep{2008A&A...487..837F}.}
\label{fig:sed}
\end{figure}

\begin{figure*}
\centering
\includegraphics[height=5.8cm]{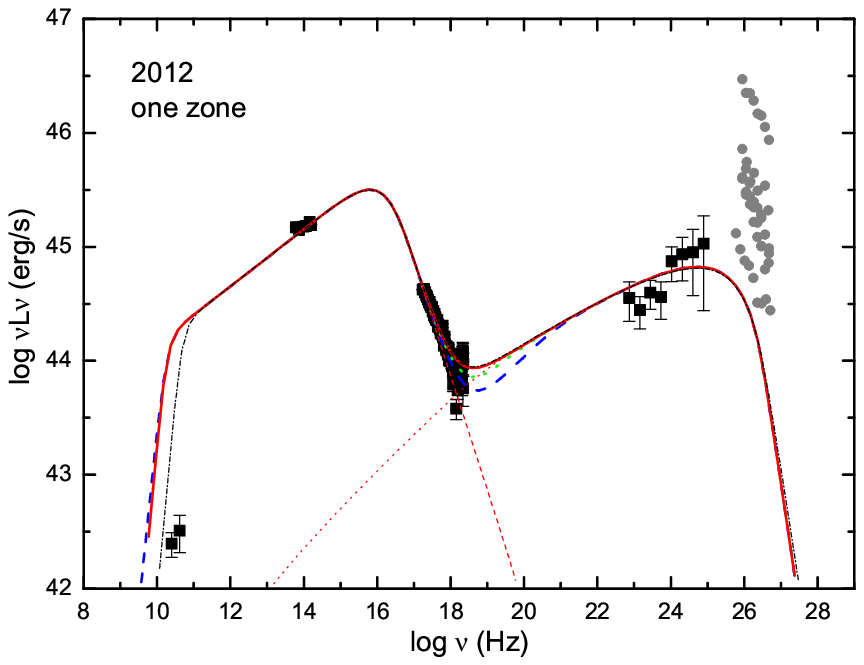}
\includegraphics[height=5.8cm]{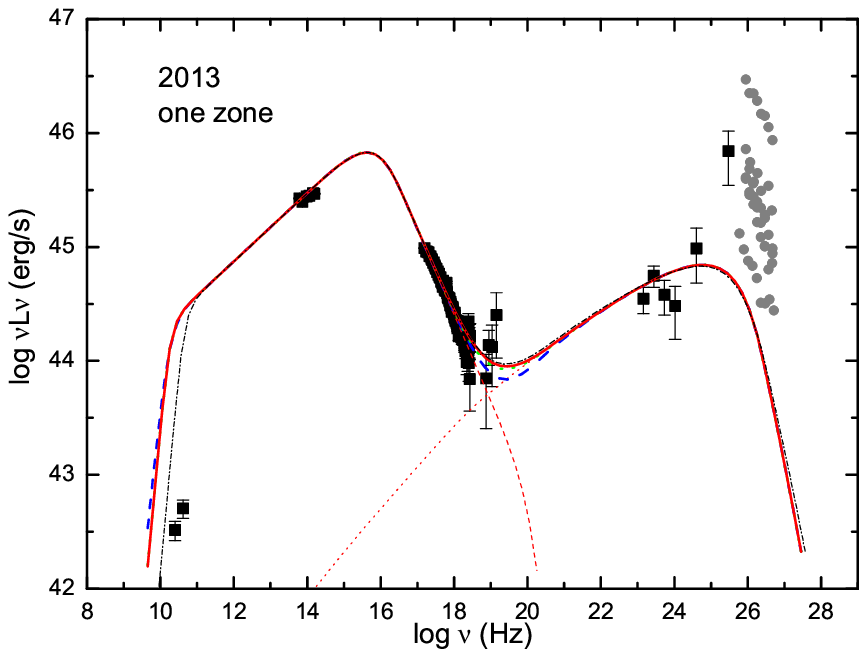}
\caption{One-zone SSC model for SED modeling of PKS 2155-304 for epochs 2012 (left panel) and 2013 (right panel).
For both epochs, the red solid lines represents the total emissions with $\gamma_{\rm min}=10$, while the red dashed
and dotted lines show the synchrotron and SSC emissions, respectively. The green dotted and blue dashed lines are
for $\gamma_{\rm min}=30$ and 100, respectively. The dash-dotted black lines are for SED modeling with $\Delta t=1$ day
 (and therefore smaller $R$ and larger $B$, see text for details). It can be seen that if one tries to fit the hard X-ray excess within
the one-zone SSC model, the model will predict larger radio fluxes than the observed ones for both epochs.}
\label{fig:sedonezone}
\end{figure*}

\begin{figure*}
\centering
\includegraphics[height=5.8cm]{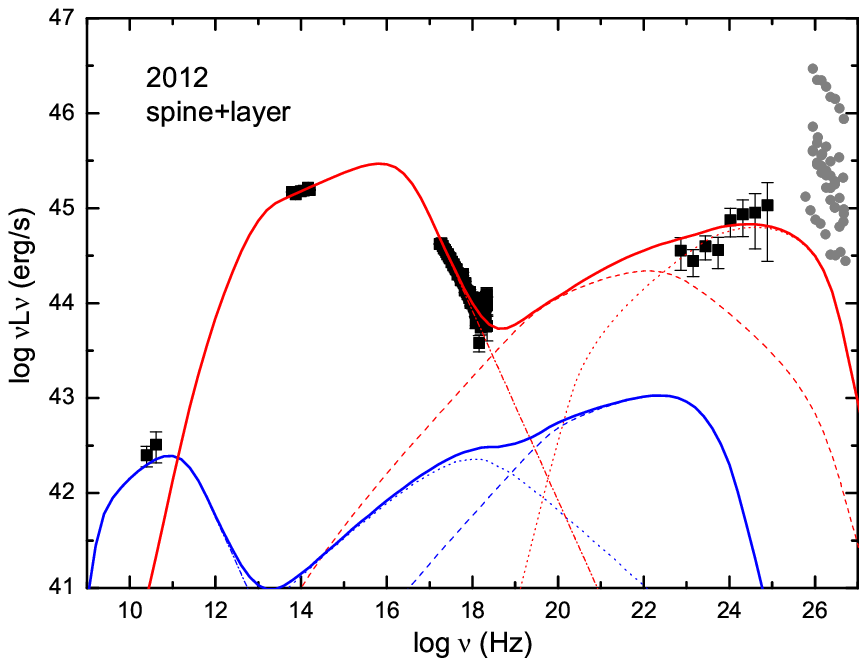}
\includegraphics[height=5.8cm]{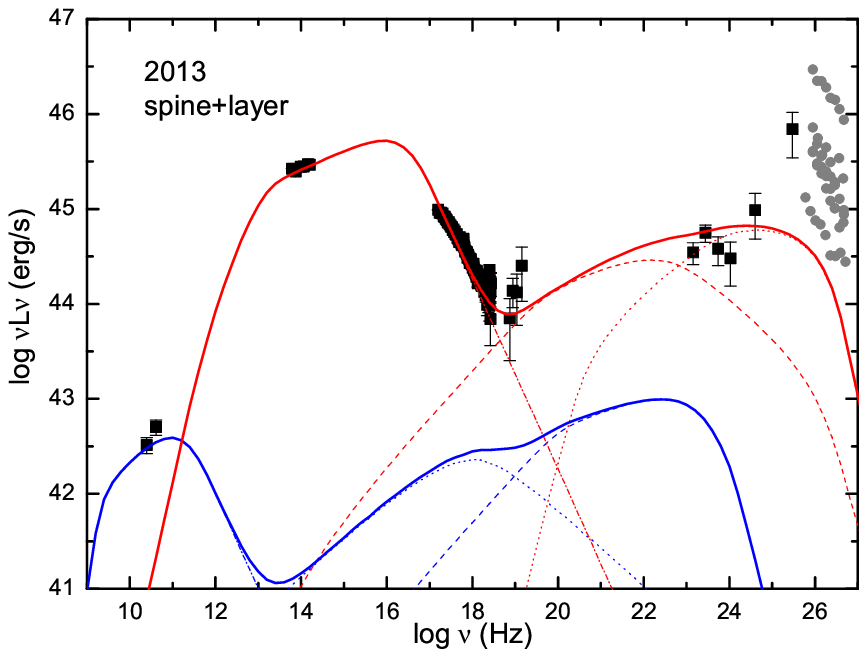}
\caption{Spine/layer model for SED modeling of PKS 2155-304 for epochs 2012 (left panel) and 2013 (right panel).
For both epochs, the red and blue lines represent the spine and layer emissions, respectively. While the dot-dashed
lines are the synchrotron emissions, the dotted lines show the SSC emissions, and the dashed lines are for the IC
emissions of seed photons originated externally from the layer/spine.}
\label{fig:sedspinelayer}
\end{figure*}

{

\appendix

\section{Data Analysis Techniques}

\subsection{Excess Variance and Variability Amplitude}

Each light curve has finite uncertainties $\sigma_{err}$ due to the measurement errors which contribute to
an additional variance. In order to subtract the uncertainties due to the individual flux measurements we calculate
the `excess variance' (Nandra et al.\ 1997; Edelson et al.\ 2002; Vaughan et al.\ 2003), which is an estimator of the
intrinsic source variance. The variance, after subtracting the excess contribution from the measurement errors is given as
(Vaughan et al. 2003)
\begin{equation}
\sigma_{\rm{NXS}}^{2} = S^{2} - \overline{\sigma_{\rm{err}}^2},
\end{equation}
where $\overline{\sigma_{\rm{err}}^{2}}$ is the mean square error,
\begin{equation}
\overline{\sigma_{\rm{err}}^{2}} = \frac{1}{N}\sum_{i=1}^{N} \sigma_{{\rm
err}, i}^{2}.
\end{equation}
The normalized excess variance is given by
$\sigma_{\rm{NXS}}^{2}=\sigma_{\rm{XS}}^{2}/\bar{x}^{2}$ and the fractional root mean square (rms) variability
amplitude ($F_{\rm{var}}$;
Edelson, Pike \& Krolik 1990; Rodriguez-Pascual et al.\ 1997) is
\begin{equation}
F_{\rm{var}} = \sqrt{ \frac{S^{2} -
\overline{\sigma_{\rm{err}}^{2}}}{\bar{x}^{2}}}.
\end{equation}
and the error on the fractional amplitude is given as
\begin{eqnarray}
{err(F_{\rm{var}}) =
\frac{1}{2 F_{\rm{var}}} err(\sigma_{\rm{NXS}}^{2}) = }
\nonumber \\
{\qquad
\sqrt{ \left\{ \sqrt{\frac{1}{2N}} \cdot\frac{ \overline{\sigma_{\rm{err}}^{2}}
}{  \bar{x}^{2}F_{\rm{var}} }  \right\}^{2}
+
\left\{ \sqrt{\frac{\overline{\sigma_{\rm{err}}^{2}}}{N}}
\cdot\frac{1}{\bar{x}}  \right\}^{2}  }.
}
\end{eqnarray}

We calculated $F_{\rm{var}}$ for all of our light curves and the results are presented in Table 1.

}
\end{document}